\documentclass[pra,twocolumn,showpacs,preprintnumbers,amsmath,amssymb,floatfix,superscriptaddress]{revtex4}
\usepackage{epsfig}
\usepackage{graphicx}
\usepackage{amssymb}
\usepackage{color}
\usepackage{lipsum}
\usepackage{verbatim}
\usepackage[normalem]{ulem}

\begin{document}

\title{Real-time Dynamics in U(1) Lattice Gauge Theories with Tensor Networks}
\author{T. Pichler}
\affiliation{Institute for complex quantum systems \& Center for Integrated Quantum Science and Technologies, Universit\"at Ulm, D-89069 Ulm, Germany}
\author{M. Dalmonte}
\affiliation{Institute for Theoretical Physics, University of Innsbruck, A-6020 Innsbruck, Austria}
\affiliation{Institute for Quantum Optics and Quantum Information of the Austrian Academy of Sciences, A-6020 Innsbruck, Austria}
\author{E. Rico}
\affiliation{Department of Physical Chemistry, University of the Basque Country UPV/EHU, Apartado 644, 48080 Bilbao, Spain}
\affiliation{IKERBASQUE, Basque Foundation for Science, Maria Diaz de Haro 3, 48013 Bilbao, Spain}
\affiliation{IPCMS (UMR 7504) and ISIS (UMR 7006), University of Strasbourg and CNRS, 67000 Strasbourg, France}
\author{P. Zoller}
\affiliation{Institute for Theoretical Physics, University of Innsbruck, A-6020 Innsbruck, Austria}
\affiliation{Institute for Quantum Optics and Quantum Information of the Austrian Academy of Sciences, A-6020 Innsbruck, Austria}
\author{S. Montangero}
\affiliation{Institute for complex quantum systems \& Center for Integrated Quantum Science and Technologies, Universit\"at Ulm, D-89069 Ulm, Germany}
\date{\today}
\begin{abstract}
Tensor network algorithms provide a suitable route for tackling real-time dependent problems in lattice gauge theories, enabling the investigation of out-of-equilibrium dynamics. We analyze a U(1) lattice gauge theory in (1+1) dimensions in the presence of dynamical matter for different mass and electric field couplings, a theory akin to quantum-electrodynamics in one-dimension, which displays string-breaking: the confining string between charges can spontaneously break during quench experiments, giving rise to charge-anticharge pairs according to the Schwinger mechanism. We study the real-time spreading of excitations in the system by means of electric field and particle fluctuations: we determine a dynamical state diagram for string breaking and quantitatively evaluate the time-scales for mass production. We also show that the time evolution of the quantum correlations can be detected via bipartite von Neumann entropies, thus demonstrating that the Schwinger mechanism is tightly linked to entanglement spreading. To present the variety of possible applications of this simulation platform, we show how one could follow the real-time scattering processes between mesons and the creation of entanglement during scattering processes. Finally, we test the quality of quantum simulations of these dynamics, quantifying the role of possible imperfections in cold atoms, trapped ions, and superconducting circuit systems. Our results demonstrate how entanglement properties can be used to deepen our understanding of basic phenomena in the real-time dynamics of gauge theories such as string breaking and collisions.
\end{abstract}



\maketitle

\section{Introduction}

The mechanism of quark confinement stands as a key concept in our understanding of the fundamental interactions in high energy physics~\cite{Halzen,Degrand,Langacker,Rothe}. As a quark and an anti-quark are pulled apart, the energy stored in the gluon string connecting them grows linearly with distance, due to the confining nature of strong nuclear forces described by quantum-chromodynamics (QCD). In gauge theories hosting dynamical charges, there exists a critical length scale at which the confining string breaks, creating particle-antiparticle pairs which reduce the energy density in the string and give rise to the hadrons at the string edges~\cite{Bali}. 

The {\it static properties} of string breaking have been widely explored using a variety of lattice methods, wherein the effective string potential separating static charges can be extracted by the Polyakov or Wilson loops \cite{Philipsen, Knechtli, Bali2}. However, the {\it real-time dynamics} of gauge theories are usually biased by a severe sign problem, and as such cannot be accessed using lattice Montecarlo simulations~\cite{Wilson,Kogut1,Kogut2}. In this paper, we apply tensor network (TN) methods in order to study the real-time dynamics of a lattice gauge theory (LGT) with dynamical charges and quantum gauge degrees of freedom in one dimensional systems. In particular, we investigate the real-time string-breaking dynamics in Abelian U(1) LGTs in (1+1)d, which share with QCD the basic feature of confinement.

In recent years, efficient numerical methods based on TNs have found widespread application to the real-time dynamics of strongly correlated low-dimensional systems~\cite{Schollwoeck1}. They are nowadays routinely used to tackle a variety of condensed matter and atomic physics problems, such as the evaluation of spectral functions of low-dimensional magnets and the quench or controlled dynamics of ultra-cold quantum gases in optical lattices~\cite{White,Fannes,Dukelsky,Verstraete1,Schollwoeck2,Perez,Chiara,Verstraete2,Murg,osterloh,monta09,sanz,gerster,Silvi1}. While TN methods have been extensively applied to spin and Hubbard-type models, only recently it has been shown how TNs can provide an ideal platform for the investigation of gauge theories, for example in the study of the static properties of the Schwinger model, the low-energy mass excitation spectrum, and the dynamics of deconfinement in 2D $\mathbb{Z}_2$ LGTs~\cite{Byrnes,casanova,Banuls1,Banuls2,Rico,Buyens,Kuhn,Silvi2,Tagliacozzo1,Haegeman,garcia}. In particular, the quantum link model (QLM) formulation of LGT~\cite{Horn,Orland,Chandrasekharan} -- gauge theories whose link Hilbert space is finite-dimensional -- has been used to develop efficient general-purpose TN algorithms to describe static and real-time dynamical properties of Abelian and non-Abelian LGTs including generic forms of matter fields, and to present different possible quantum simulator implementations on different platforms: trapped ions, cold atoms in optical lattices and circuit quantum electrodynamics (QED)~\cite{Weimer,Kapit,Zohar0,Zohar1,Banerjee,Zohar2,Marcos1,Hauke,Tagliacozzo2,Zohar3,Marcos2,Stannigel,Glaetzle,Notarnicola,Meurice} (see also~\cite{WieseReview,ZoharReview} for recent reviews and references therein). 

\begin{figure*}[t]
\includegraphics[width = 1.99\columnwidth]{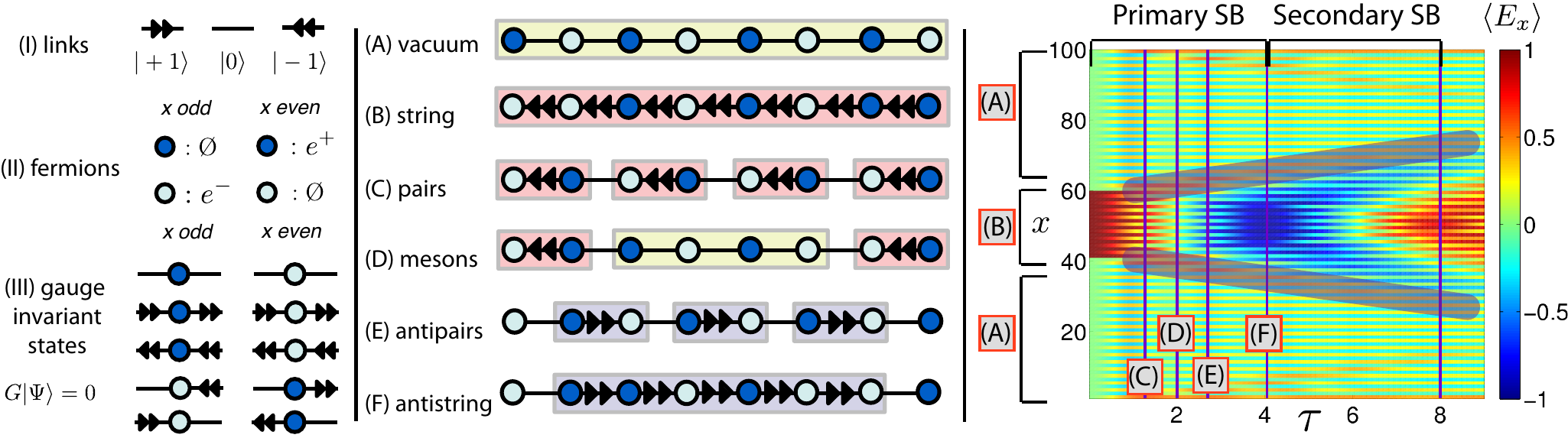}
\caption{Left panel: Hilbert space and gauge invariant states of the QLM. {\it (i)} In the quantum link formulation, the gauge fields defined on the links are described by spins (in our case, S=1). {\it (ii)} Staggered fermions represent matter and antimatter fields on a lattice bipartition: on the even (odd) bipartition, a full (empty) site represents a particle (antiparticle). {\it (iii)} Hilbert space and gauge invariant states of the QLM. The Gauss law, Eq. (\ref{eq:GaussLaw}), constrains the number of possible states at each lattice site. Notice that the Gauss law depends on the lattice site due to the staggered fermions. Middle panel: cartoon states for the different stages of the string breaking dynamics (see text). Here the leftmost site is an odd one. Right panel: sample simulation for the electric field dynamics when quenching an initial string state (B region) connecting two charges, and surrounded by the vacuum (A regions) for $m=0=g$. Primary string breaking takes place in four stages (C-F), until an anti-string is created in place of the original string. The latter also decays during the secondary string breaking. The shaded areas represent the wave-fronts estimated from entanglement entropies (see Sec.~\ref{sec:Ent}), which are directly related to the electric field evolution. }
\label{summary:fig}
\end{figure*}

TN methods are based on variational tensor structure ansatze for the many-body wave function of the quantum system of interest: the tensor structure is chosen to best accommodate some general system properties, e.g., dimensionality, boundary conditions and symmetries, while a controlled approximation is introduced in such a way that one can interpolate between a mean field and an exact representation of the system. Being a wave function based method, one has direct access to all relevant information of the system itself, including quantum correlations, i.e., entanglement. In one-dimensional systems, an efficient tensor structure is given by the Matrix Product State (MPS) ansatz \cite{Fannes,Schollwoeck1}, defined as, 
\begin{equation}
|\psi_{\mathrm MPS}\rangle = \sum_{\vec \alpha} A_{\alpha_1}^{\beta_1} A_{\alpha_2}^{\beta_1,\beta_2} \dots A_{\alpha_N}^{\beta_{N-1}} |\vec \alpha \rangle,
\label{MPSeq}
\end{equation}
where the tensor $A$ contains the variational parameters needed to describe the system wave-function, $\alpha_i=1, \dots, d$ characterize the local Hilbert space, and $\beta_i=1, \dots, m$ account for quantum correlations or entanglement (Schmidt rank) between different bipartitions of the lattice. Indeed, setting $m =1$ results in a mean field description, while any $m >1$ allows for the description of correlated many-body states. Given the tensor structure, the tensor dimensions and coefficients are then optimized to efficiently and accurately describe the system properties by means of algorithms that scale polynomially in the system size and $m$. Usually, these algorithms exploit the system Hamiltonian tensor structure, naturally arising from the few-body and local nature of the interactions, to efficiently describe the system ground state or low-lying eigenstates, or to follow the real- or imaginary-time evolution of the system itself. Indeed, in the TN approach, real- and imaginary-time evolution have no fundamental differences at the computational level as there is no sign problem, and limitations arise -- only in some scenarios depending on the specific dynamics of interest 
as witnessed by the fast increasing literature appearing based on this approach~\cite{Schollwoeck1} -- from the amount of quantum correlations present in the system wave function.

Here, we show how TN algorithms allow for the study of the real-time dynamics of LGTs, focusing on the string breaking in a paradigmatic confining theory - the Schwinger model \cite{Schwinger0,Schwinger1,Schwinger2} in a quantum link formulation. We characterize the real-time dynamics of the primary and secondary string breaking and show that string breaking is intimately related to entanglement production in the system. A qualitative picture for string breaking in our models, together with a typical result for our time-dependent simulations on a system of $L=100$ lattice sites, is illustrated in Fig.~\ref{summary:fig}. Even more importantly, our simulations allow us to track the entanglement evolution during string breaking: as we will show below, string breaking and the so-called Schwinger mechanism are intimately connected to entanglement propagation, which we address by evaluating the so called von Neumann entanglement entropy. Finally, we show that TN methods can be used to study scattering processes between bound states of LGTs: we develop a scheme to engineer {\it meson} collisions~\cite{Feng}, and we show how, very surprisingly, the scattering not only reflects into an enhanced rate of particle-antiparticle creation, but it does affect drastically the entanglement properties of the system, which stays significantly correlated well beyond the scattering time-window. 

The paper is structured as follows: in Sec.~\ref{sec_model} we present the system Hamiltonian and recall the TN algorithm we are using throughout this work. In Sec.~\ref{sec:StrBreaking} we present the results on string breaking and mass production dynamics, including a discussion on how this phenomenon can be observed using a quantum simulation platform. In Sec.~\ref{sec:Ent} we show how entanglement follows the string breaking dynamics, providing a quantitative picture which underlines how entanglement entropies are directly tied to string-breaking. Finally, we present our result on scattering in Sec.~\ref{sec:Scatt}, and draw a summary of our results in Sec.~\ref{Sec:Concl}.

\section{Model and Methods}\label{sec_model}

\subsection{Model Hamiltonian: QED in (1+1)d}

QED in (1+1)d, also known as the Schwinger model~\cite{Schwinger0}, represents an ideal testing-ground for the benchmarking and development of new computational methods. Despite its relative simplicity, this model captures fundamental aspects of gauge theories such as, e.g., the presence of a chiral symmetry undergoing spontaneous symmetry breaking~\cite{Schwinger0,Schwinger1,Schwinger2,Lowenstein,Casher,Coleman1,Polyakov,Banks,Drell,Ben,Coleman2}. Even more importantly, this theory, like QCD, displays confinement: differently from (3+1)d QED, in (1+1)d electrons and positrons are confined, and interact via a long-range potential which increases linearly with distance. Due to the large energy cost associated with the electric flux between charges at large inter-charge distances, the electric flux string is unstable to particle-antiparticle creation, as in QCD, and string breaking takes place~\cite{Hebenstreit}. While this phenomenon, directly connected to the Schwinger mechanism of mass production out of a vacuum, has long been debated, and notable insights have been provided using a variety of approximate methods, a full quantum mechanical understanding of the complex real-time dynamics taking place during string breaking is lacking due to the computationally complexity of the many-body problem~\cite{Szpak,Hebenstreit,Kasper,BergesReview}. 

In the Hamiltonian formulation, its dynamics are defined by the following form:
\begin{eqnarray}
 H&=&-t\sum_x \left[ \psi^{\dagger}_x U_{x,x+1}^{\dagger} \psi_{x+1} + \psi^{\dagger}_{x+1} U_{x,x+1} \psi_{x} \right] \nonumber\\
  & &+m\sum_x (-1)^x\psi^{\dagger}_x\psi_x
  +\frac{g^2}{2}\sum_xE^2_{x,x+1}.
   \label{eq:Hamiltonian}
\end{eqnarray}
where $\psi^\dagger_x,\psi_x$ are fermionic creation/annihilation operators describing Kogut-Susskind (staggered) fermions (see Fig.~\ref{summary:fig}), $U_{x,x+1}$ are the gauge fields residing on the $(x,x+1)$ link, and we denote the strength of fermion-hopping (the kinetic energy of electrons and positrons) with $t$, the staggered mass of the fermions with $m$, and the electric coupling strength with $g$, where $E_{x,x+1}$ is the electric-field operator. The gauge generator is given by
\begin{eqnarray}
 \tilde{G}_x&=&\psi^{\dagger}_x\psi_x+E_{x,x+1}-E_{x-1,x}+\frac{(-1)^x-1}{2},
   \label{eq:GaussLaw}
\end{eqnarray}
so that $[H, G_x]=0$ and all physical states $|\Psi\rangle$ satisfy the Gauss law $\tilde{G}_x |\Psi\rangle=0$. While in the Wilson formulation $U_{x,x+1}$ are parallel transporters acting on an infinite dimensional Hilbert space, we focus here on a formulation based on QLMs, where the gauge fields are represented by spin-1 operators, $U_{x,x+1}=S^+_{x,x+1}, E_{x,x+1}=S^z_{x,x+1}$ and, as such, act on a finite-dimensional link Hilbert space~\cite{Banerjee}. In particular, the electric field operator allows three possible states for the electric flux, constraining the physical states per site as described in Fig.~\ref{summary:fig}. A detailed discussion of the quantum link formulation can be found in Ref.~\cite{Horn,Orland,Chandrasekharan}, while in Ref.~\cite{Rico} it was shown how such quantum link formulation reproduces the phase diagram and quantum criticality of the continuum theory. 

\subsection{String breaking and classical cartoon states}

String breaking is the process of cutting and shortening the electric flux string that connects a particle-antiparticle pair by creating a new charge-anticharge pair~\cite{Banerjee}. Within our framework, a string consists of two charges creating non-zero electric flux between them. The charges are represented by appropriate boundary conditions (static charges) or by excitations of the mass field at the site of the fermion (dynamical charges). This is realized by an effective jump of a fermion from the site of one charge to the site of the second charge satisfying the Gauss law. The string of electric flux then follows from Gauss' law. The charges force the links into a non-zero flux state, according to the configuration of the charges in either one direction or the other. Before embarking in a full quantum mechanical investigation of string breaking, we now discuss the classical ($t=0$) static picture which provides a simple yet informative illustration of the different stages of the string breaking mechanism. A set of cartoons of the classical states is provided in Fig.~\ref{summary:fig}:

{\it Vacuum.-} In the vacuum (A), neither mass nor electric field excitations are present. The vacuum energy is thus $E_0=-\frac{L}{2}m$.

{\it String.-} In the string state (B), two mass excitations are present at the boundaries, and all electric fields connecting the two are also in the $|+1\rangle$ state. The resulting string energy then takes the form
\begin{equation}
E_{\mathrm{string}}-E_0=\frac{g^2}{2}(L-1)+2m.
\end{equation}

{\it Pairs.-} In the pairs state (C) all the masses are excited forming charge-anticharge pairs with an energy $E_{\mathrm{pairs}}=\frac{g^2L}{4}+mL$.

{\it Mesons.-} In a confined phase, particle-antiparticle pair production can favor the establishment of a vacuum state between two static charges, with a pair of mesons at the boundary of the string (see (D)). The resulting energy is:
\begin{equation}
E_{\mathrm{mesons}}-E_0= g^2+4m.
\end{equation}
At the static level, string breaking takes place at a critical distance $L_c$, above which the meson state is energetically favored over the string state ($E_{\mathrm{string}}(L_c)=E_{\mathrm{mesons}}$):
\begin{equation}
L_c = \frac{4m}{g^2} + 3.
\end{equation}

{\it Antipairs and Antistring.-} The antipair-state (E) and the antistring (F) denote the pair-state and the string with the electric flux having the opposite sign.

At a dynamical level, string breaking takes place as a consequence of the Schwinger mechanism: in (1+1)d, the vacuum between two charges of opposite sign is unstable against particle-antiparticle creation~\cite{Szpak,Hebenstreit,Kasper,BergesReview}, eventually leading to the electric field in the system flipping sign and to mass production. In real-time, this process develops following intermediate consecutive steps, and is schematically illustrated in Fig.~\ref{summary:fig}: at very short timescales, particle-antiparticle creation takes place in the middle of the string, creating the {\it pair} state depicted in (C). Subsequently, the electric field in the center of the string relaxes to 0, and the external charges get screened, effectively forming mesons (D). At this point the process reverses, first establishing a state with {\it anti-pairs} (E), which finally decays into a string of oppositely signed electric field with respect to the initial state, an {\it anti-string} (F).

String-breaking is a direct consequence of confinement: while in a deconfined phase such as QED in (3+1)d, it is possible to separate opposite charges at large distances due to Coulomb's law, in a confined phase the corresponding electric field string breaks due to the effective potential between charges increasing as a function of distance. However, the exact real-time dynamics of string breaking are inaccessible to classical simulations based on Montecarlo sampling due to a severe sign-problem. While classical-statistical approaches can provide remarkable insights in some parameter regimes~\cite{Szpak,Hebenstreit,Kasper,BergesReview} such as small masses, unbiased numerical simulations for arbitrary parameter regimes have been lacking. In the following, we present a systematic study of the string breaking dynamics using MPS techniques. 

\begin{figure}
\includegraphics[width =0.9\columnwidth]{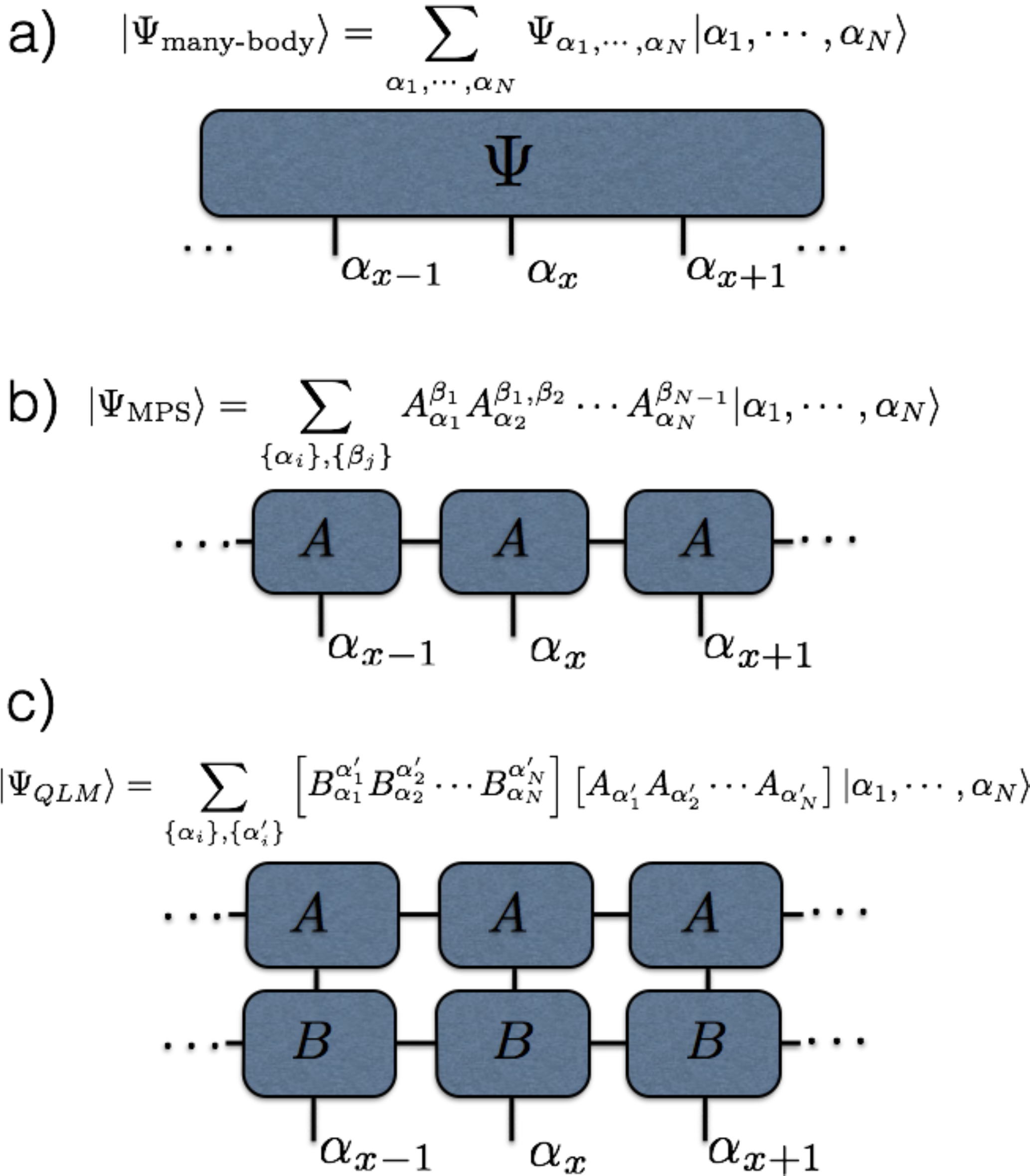}
\caption{Tensor network representations of a many-body quantum system. a) Any quantum state can be described via a tensor $\Psi$ of exponential dimension $d^{N}$, where $d$ is the dimension of the local Hilbert space and $N$ the number of sites. b) In a MPS representation, the wave function is characterized by $N$ local tensors $A$, each one of dimensions $m^{2}d$, where $m$ is the maximum Schmidt rank allowed between different bipartitions. c) Gauge-invariant tensor network: the gauge invariant state can be represented via a tensor network state where a MPO imposes the gauge invariance while a variational MPS accommodates for the detailed description of the wave function.}
\label{MPSplot}
\end{figure}

\subsection{Tensor networks for lattice gauge theories}

Tensor network algorithms are one of the paradigms for simulating quantum many-body systems in low-dimensions, both in and out of equilibrium, via a representation of the quantum state with a variational ansatz for wave-functions and/or density matrices~\cite{Schollwoeck1}.

For one-dimensional pure states, on which we focus here, the starting point is to consider a class of states of the form given in Eq.~\eqref{MPSeq} and depicted in Fig.~\ref{MPSplot} with some fixed auxiliary dimension $m$ and physical dimension $d$: for example, for spin one-half systems one has $d=2$, while the dimension $m$ depends on the states studied and on the desired accuracy. For ground states of one-dimensional gapped Hamiltonians, $m$ is in general independent of the system size $N$, while for critical systems, due to area-law logarithmic violations, the auxiliary bond dimension scales as $m \propto c \log{N}$, where $c$ is the central charge of the system (see Ref.~\onlinecite{Calabrese} for a review). The case of out-of-equilibrium dynamics as considered here is more challenging, and no general picture is known, thus the bond dimension $m$ has to be adapted to each specific case and convergence has to be checked comparing the results at increasing bond dimension~\cite{Vidal,Daley}.

Global symmetries of the system of interest, such as particle number or global magnetization and even more complex non-Abelian global symmetries, can be embedded in the wave function ansatz given in Eq.~\eqref{MPSeq} in an elegant way by promoting each tensor $A_{\alpha_j}^{\beta_j,\beta_{j+1}}$ to a symmetry-sector-preserving tensor to a tensor that preserves every symmetry sector; that is, every index of the tensor is dressed with the corresponding symmetry charge number~\cite{Ostlund,Singh0,Bauer,Weichselbaum}. 
This symmetric formulation of tensor networks allows one to describe wave functions exactly and more efficiently with the desired quantum numbers, addressing each symmetry sector separately. Recently, it has also been shown that local symmetries -- namely gauge symmetries -- can be embedded in such description by generalizing the wave function ansatz to a gauge invariant one~\cite{Rico,Buyens,Silvi2,Tagliacozzo1,Haegeman}. We briefly recall such a construction in the rest of this Section, and tailor it to the model we will study in the rest of the paper.

\begin{figure*}
\epsfig{file=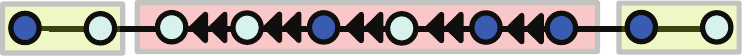,width=4.3cm,angle=90,clip=10}
\epsfig{file=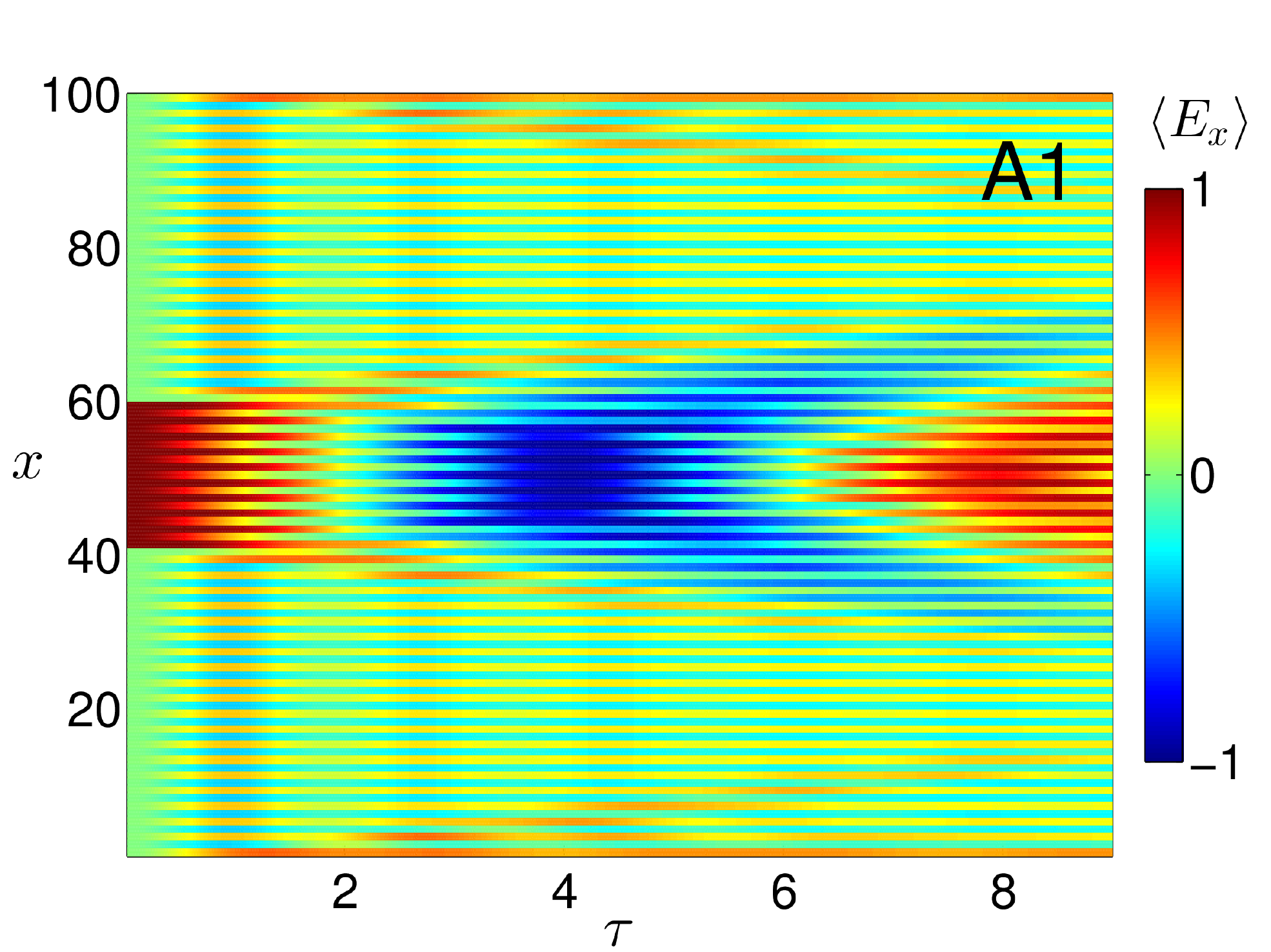,width=5.5cm,angle=0,clip=10}
\epsfig{file=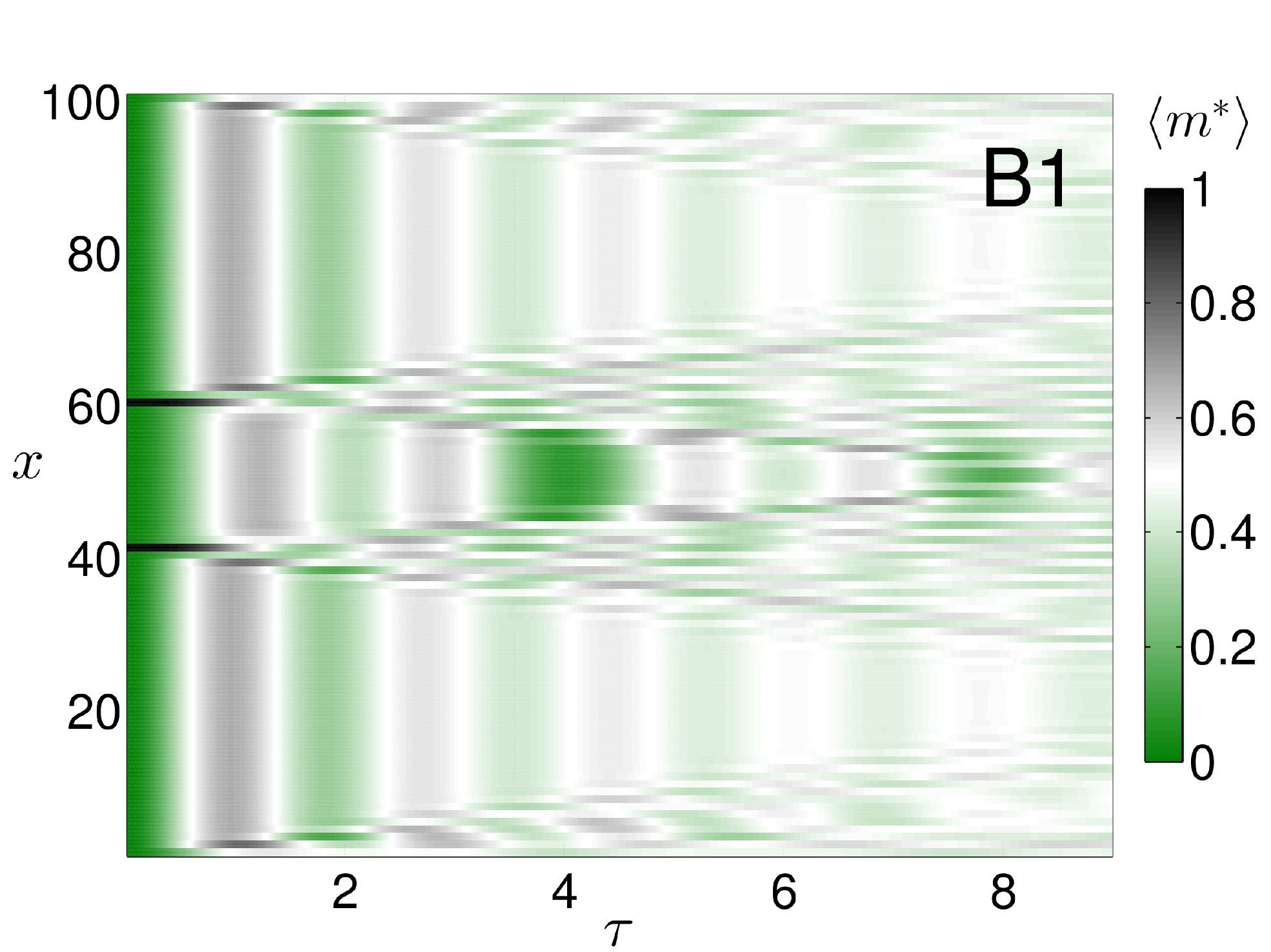,width=5.5cm,angle=0,clip=10}
\epsfig{file=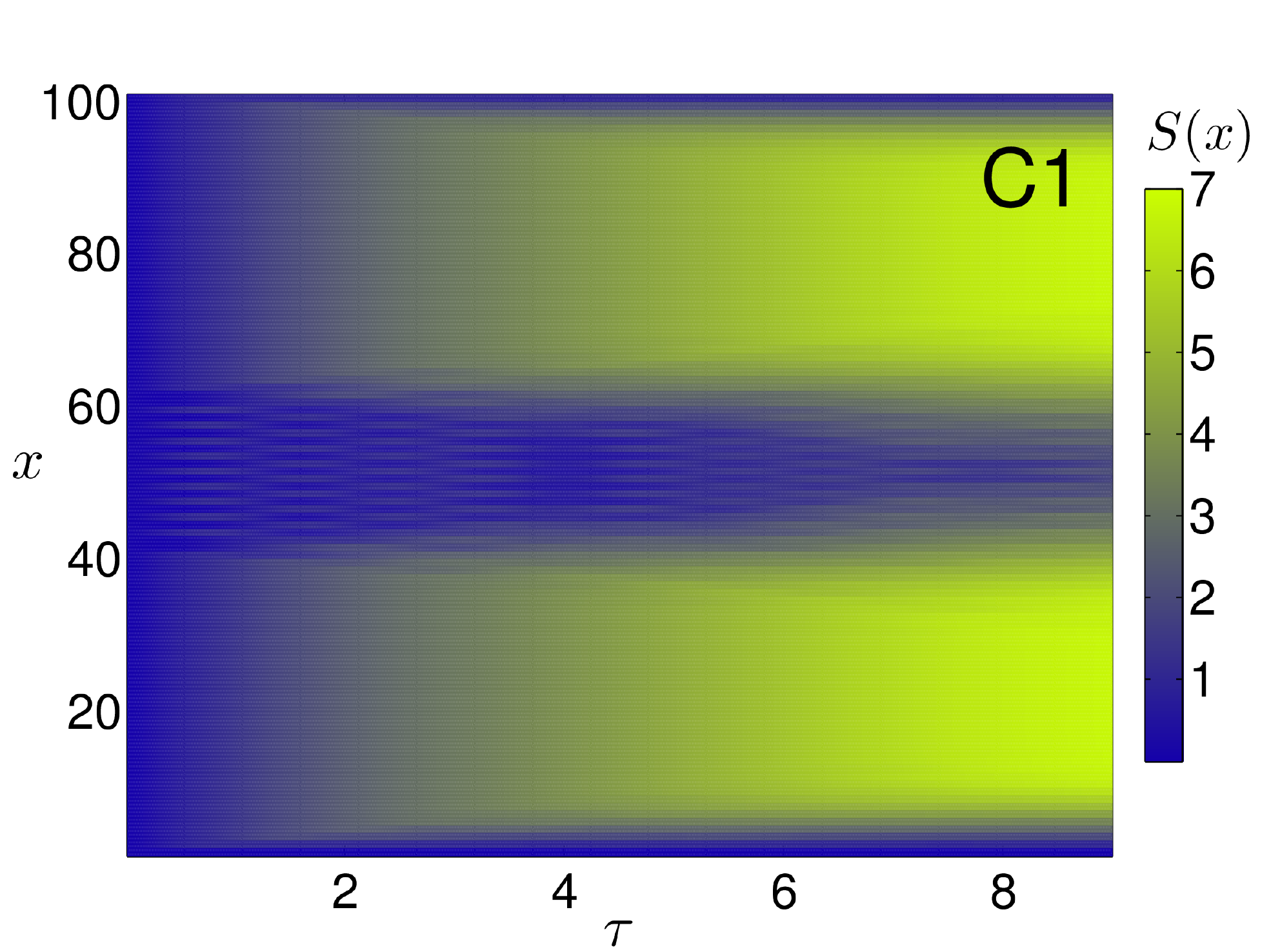,width=5.5cm,angle=0,clip=10}\\
\epsfig{file=string.pdf,width=4.3cm,angle=90,clip=10}
\epsfig{file=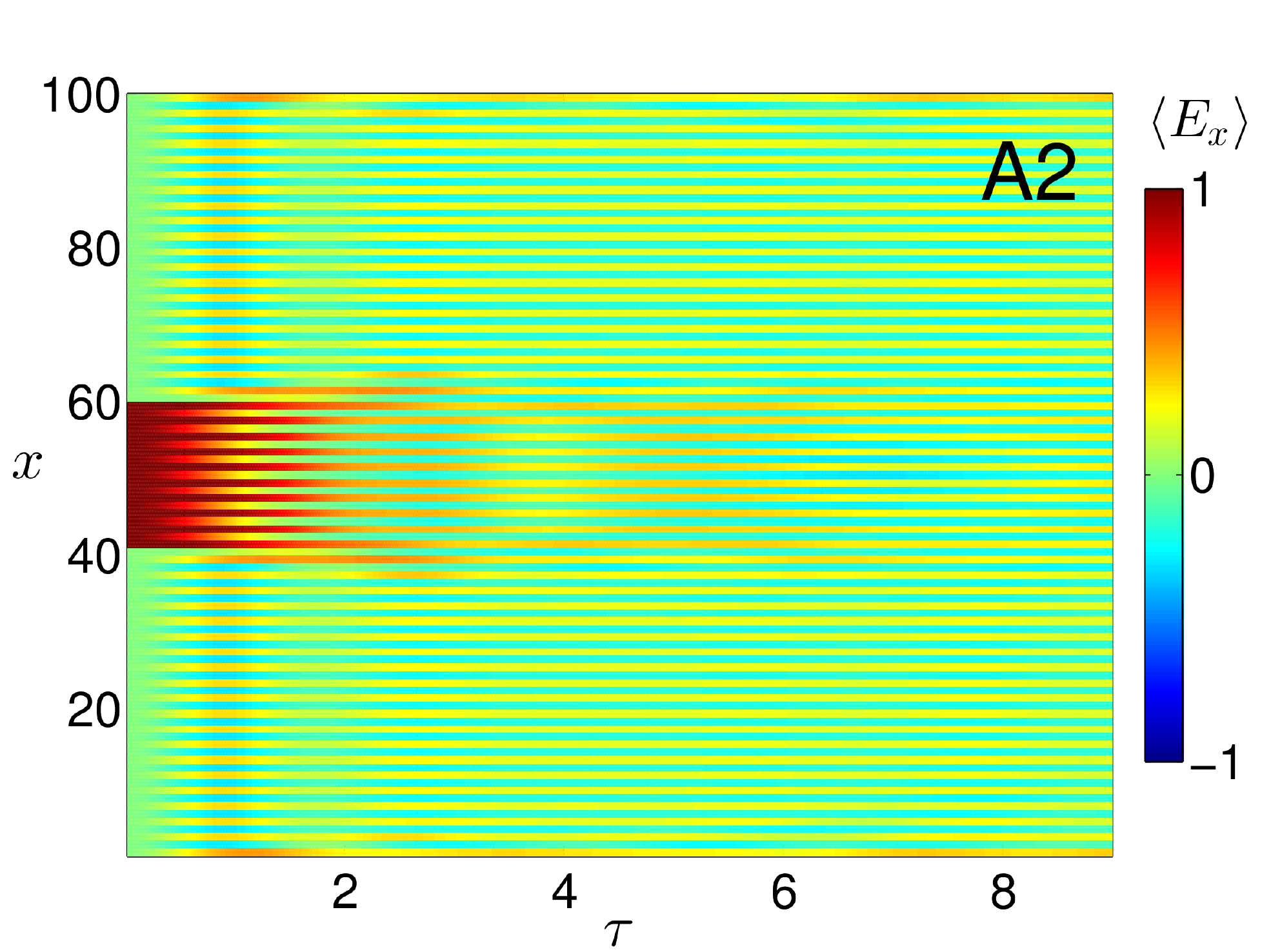,width=5.5cm,angle=0,clip=10}
\epsfig{file=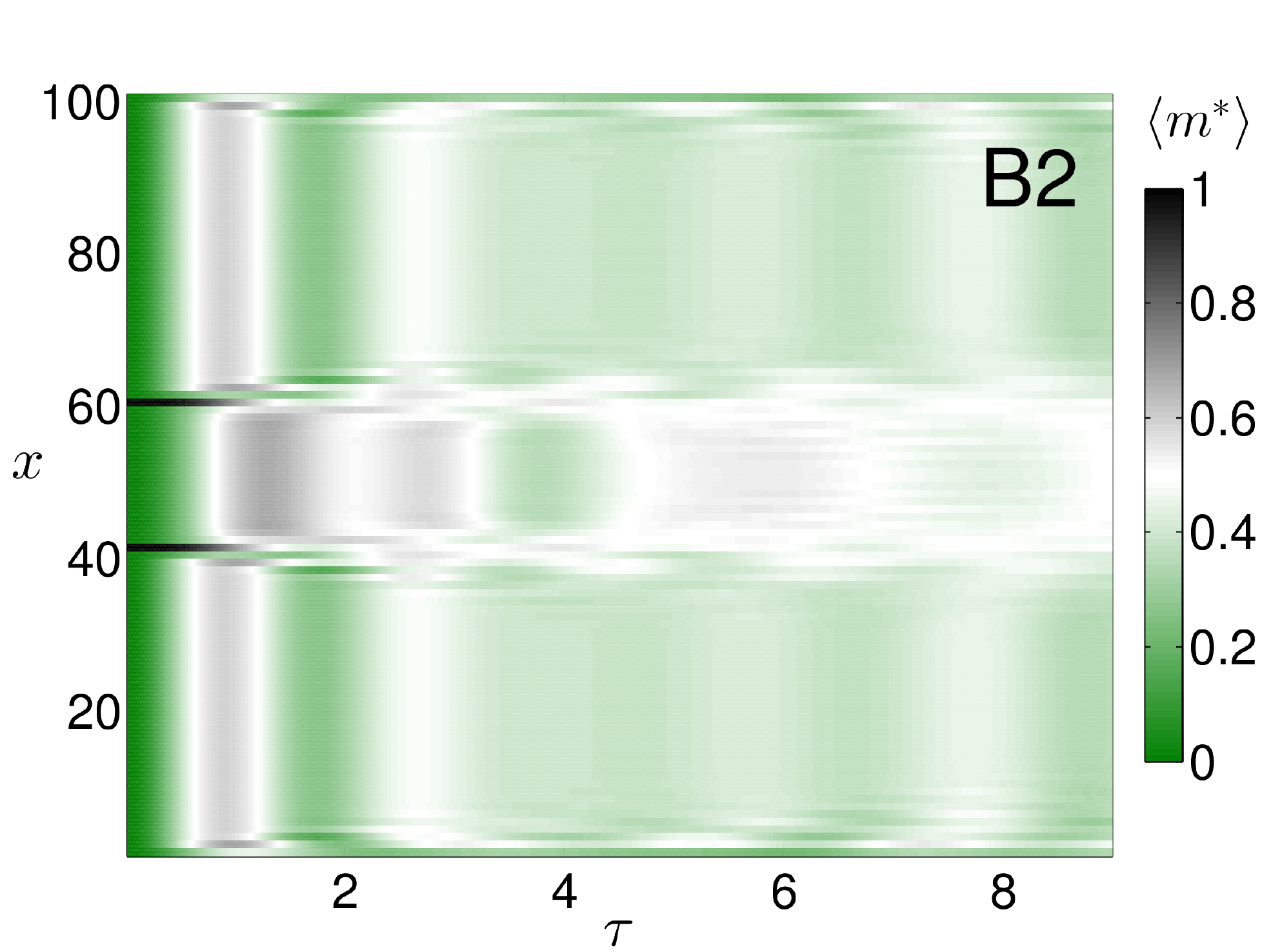,width=5.5cm,angle=0,clip=10}
\epsfig{file=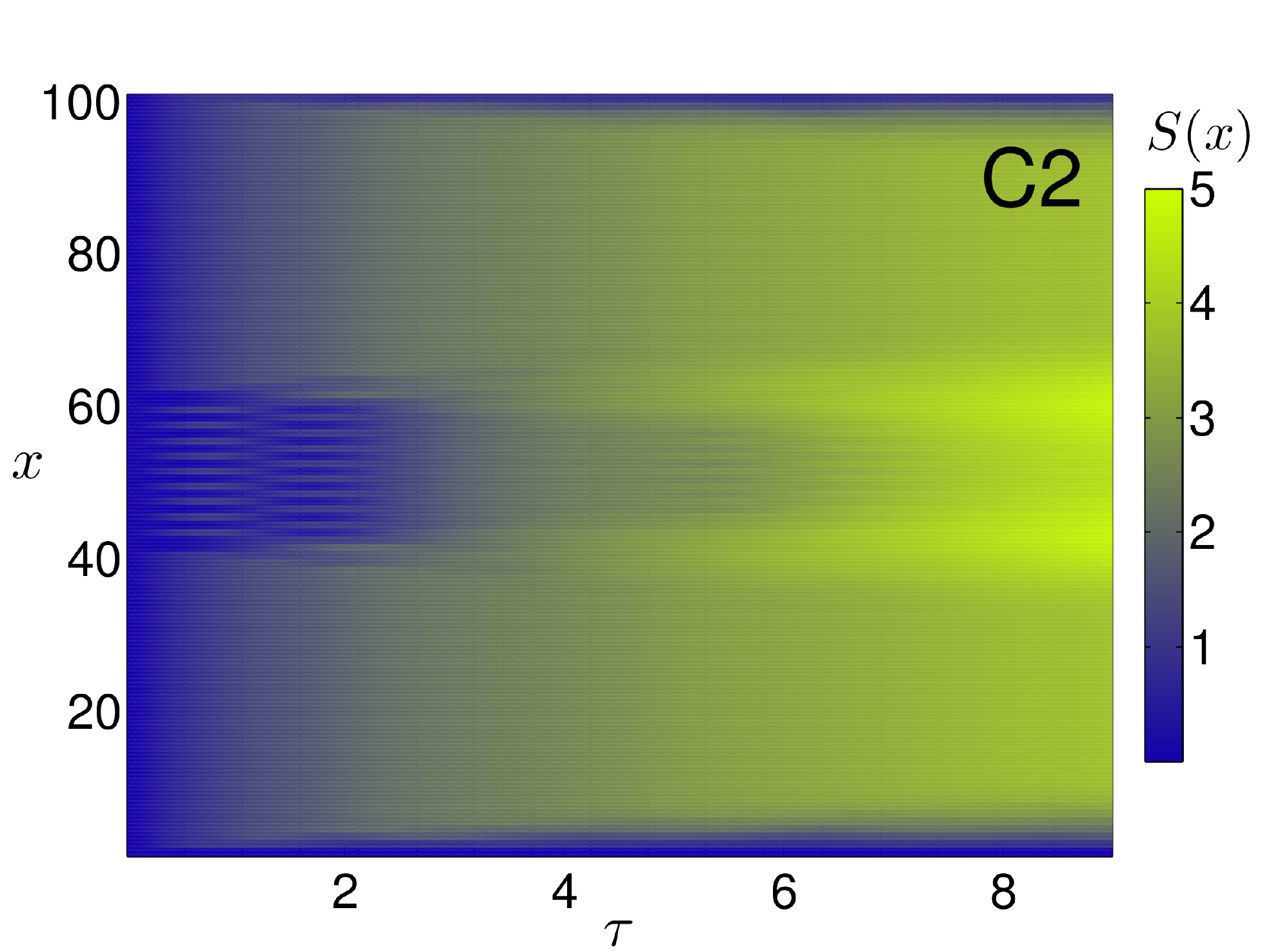,width=5.5cm,angle=0,clip=10}\\
\epsfig{file=string.pdf,width=4.3cm,angle=90,clip=10}
\epsfig{file=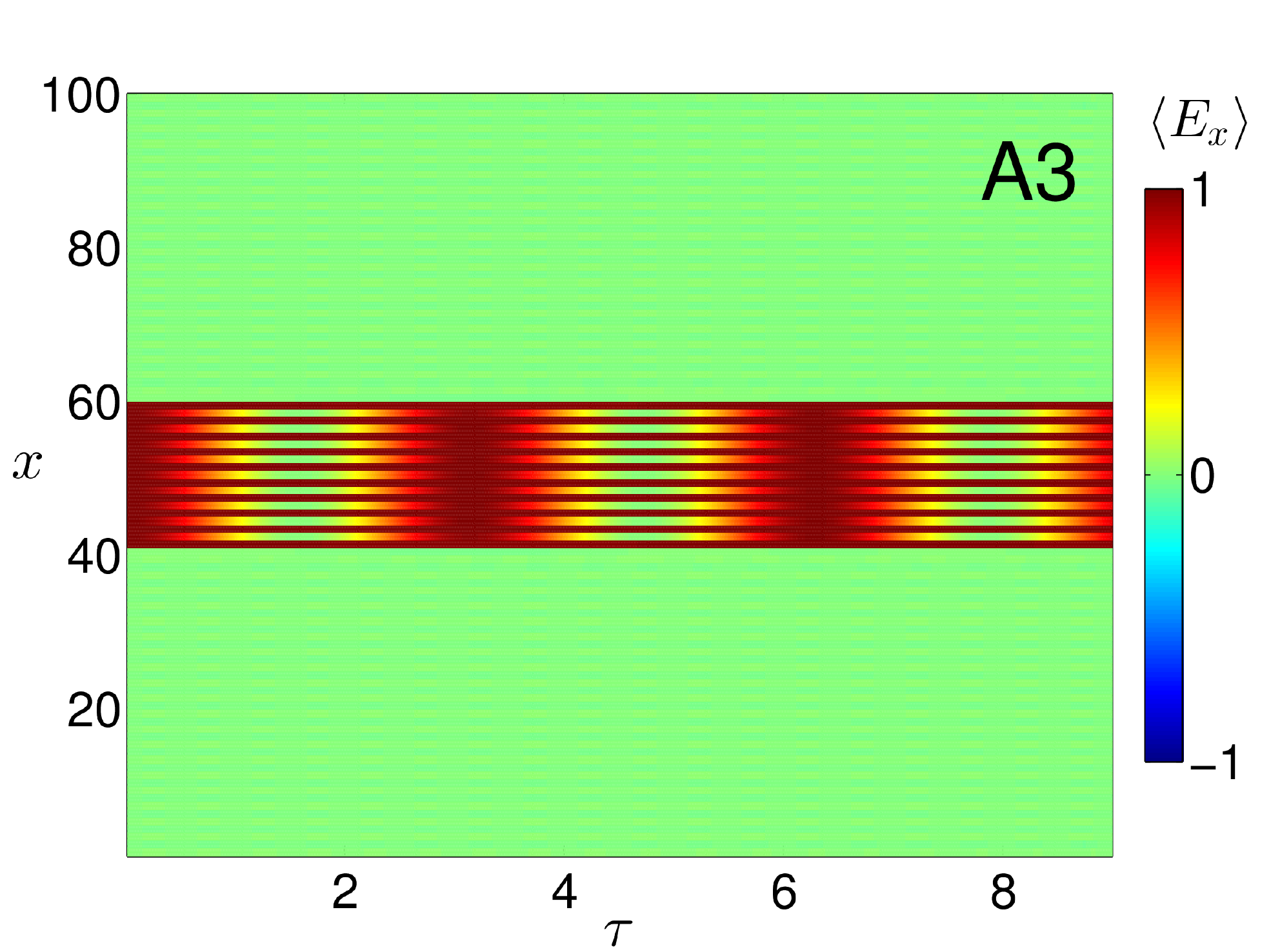,width=5.5cm,angle=0,clip=10}
\epsfig{file=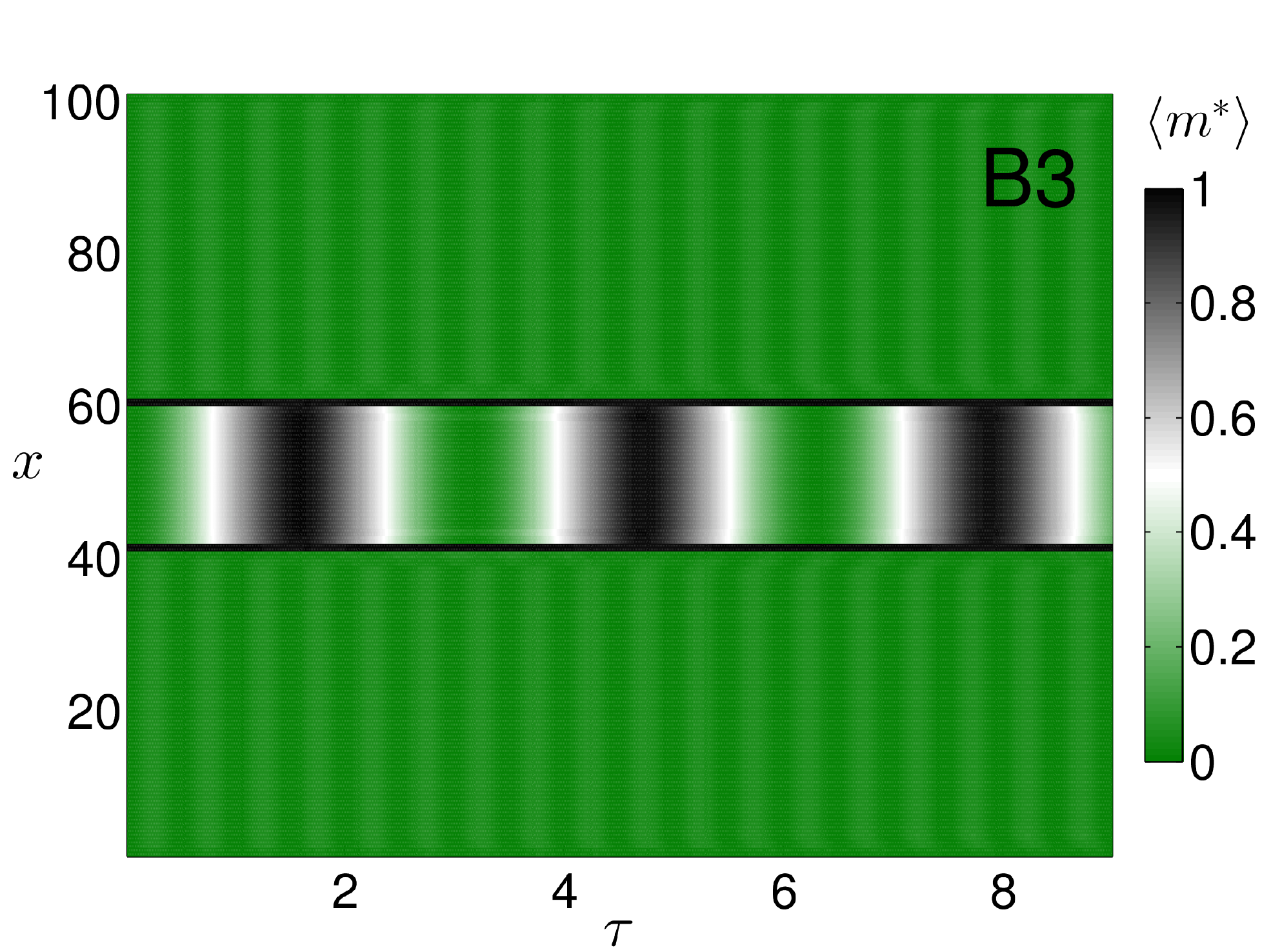,width=5.5cm,angle=0,clip=10}
\epsfig{file=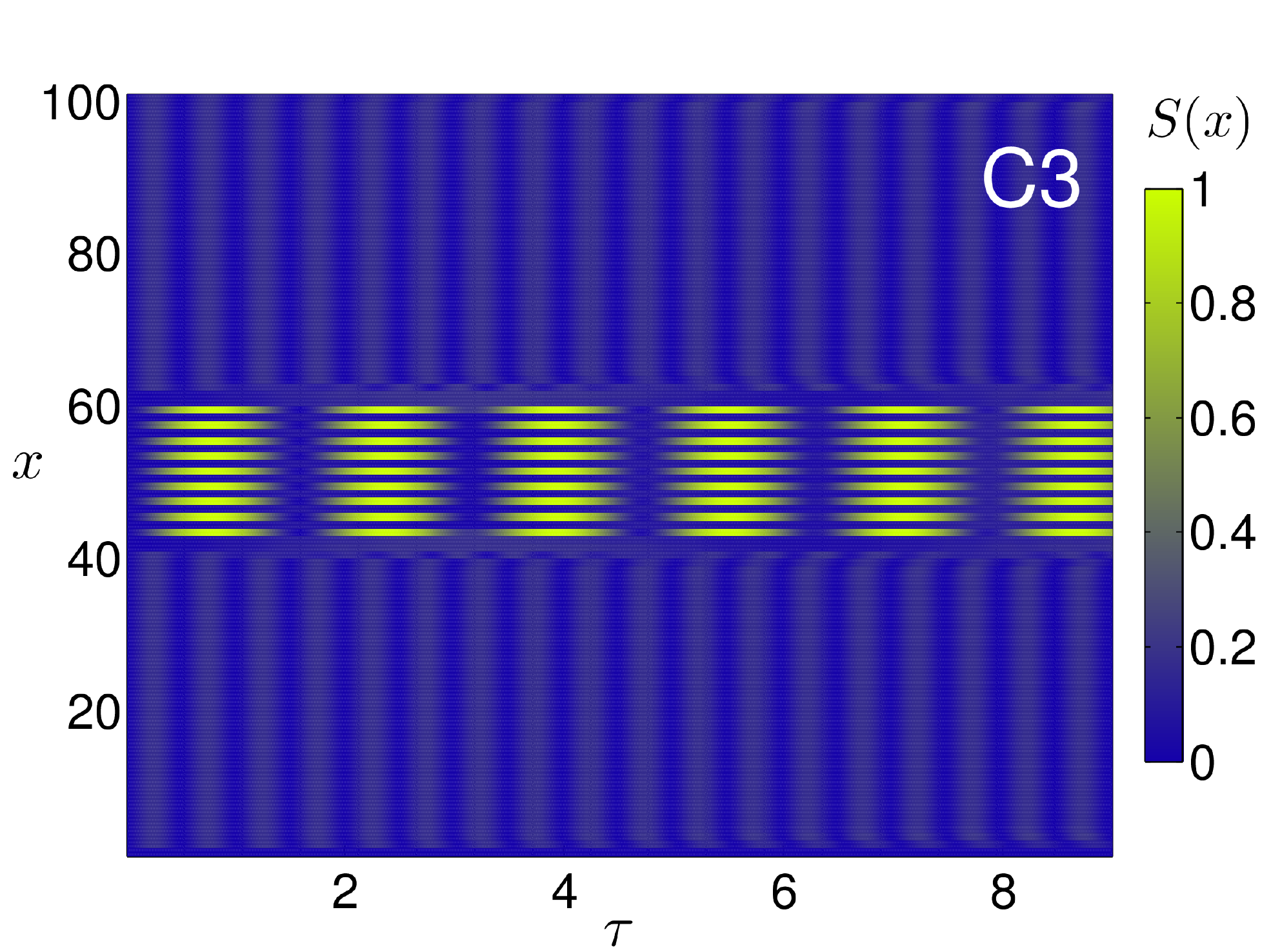,width=5.5cm,angle=0,clip=10}
\caption{Real-time evolution of a string of electric flux of length $L=20$ embedded in a larger lattice (of length $N=100$) in the vacuum state (the initial cartoon state is sketched on the left side of the figure with the notation of Fig.\ref{summary:fig}). The electric flux real-time evolution (Column A) is shown with the evolution of the mass excitations (Column B) and the evolution of the bipartite von Neumann entropy (Column C) for $m=0, g=0$ (Line 1), $m=0.25, g=1.25$ (Line 2), and $m=3, g=3.5$ (Line 3). The von Neumann entropy is calculated using a bipartition  of the system defined via a cut between lattice sites $x$ and $x+1$.}
\label{Overview:fig}
\end{figure*}

The Hilbert space of a gauge invariant system is given by the direct sum of every sector with different values of the "Gauss law'', $\tilde{G}_{x} |\Psi_{QLM} \rangle = g_{x} |\Psi_{QLM} \rangle$ and $\mathcal{H} = \oplus_{g_{x}} \mathcal{H}_{g_{x}} $. Due to the gauge symmetry, there is no physical (e.g., gauge invariant) operator, including the Hamiltonian, that connects any two different sectors. Hence, starting with a quantum state partially defined by the values of the Gauss law, or a set of local (gauge) constants of motion, the time-evolved wave-function will remain in this sector under the action of the Hamiltonian. In the QLM formulation of LGTs, a gauge invariant tensor description is immediately obtained if we use a "rishon''~\cite{Chandrasekharan} or Schwinger representation of the gauge degrees of freedom (independent of the nature of the local continuous symmetry - Abelian or non-Abelian - and the dimensionality of the lattice~\cite{Rico}).

In our case, the $U(1)$ QLM in (1+1)-d with a spin-1 per link, the spin operator or electric field $E_{x,x+1} = S^{(z)}_{x,x+1}$ is represented by a pair of Schwinger bosons $E_{x,x+1} = \frac{1}{2} \left( n_{R,x+1} - n_{L,x} \right)$ with a total occupation of two bosons per link, i.e. $n_{R,x+1} + n_{L,x} =2$. The first step to build an efficient tensor network representation of such states is to identify a local Hilbert space spanned by the states $|\alpha\rangle$ defined on the tensor product of the fermionic matter field on a lattice site and of the rishon states on its left and right, that is $|\alpha\rangle =  |n_{R,x} , n_{\Psi, x}, n_{L, x} \rangle$. This allows for the projection of the state into the gauge invariant subspace, restricting the local bases only to the "physical'' states: in our model, this process results in the five gauge invariant states (uniquely identified in terms of the Schwinger rishons for even and odd lattice sites respectively and depicted in Fig.\ref{summary:fig} left) given by~\cite{Banerjee}:
\begin{equation}
\begin{split}
&\text{odd: } |1 \rangle = |1,0,2\rangle, ~ \, |2 \rangle = |2,0,1\rangle, \\
|3 \rangle & = |1,1,1\rangle, ~ \, |4 \rangle = |2,1,0\rangle, ~ \, |5 \rangle = |0,1,2\rangle, \\
&\text{even: } |1 \rangle = |1,1,0\rangle, ~ \, |2 \rangle = |0,1,1\rangle, \\
|3 \rangle & = |1,0,1\rangle, ~ \, |4 \rangle = |2,0,0\rangle, ~ \, |5 \rangle = |0,0,2\rangle. \\
\end{split}
\end{equation}
These are the only states allowed locally by gauge invariance (notice that the local Hilbert space is different on even and odd sites as a consequence of the staggered nature of the fermions). A gauge invariant many-body state clearly exists in the tensor product of such local basis states. However, not all  tensor product combinations are compatible with the spin representation $S$: with our choices, only the states with $n_{R,x+1} + n_{L,x} =2$ are allowed. This additional constraint can be satisfied by modifying the tensor structure of the many-body state ansatz, that is by introducing a projector which acts on nearest-neighbor lattice sites and enforces the correct number of rishons on the link~\cite{Rico,Silvi2}.

Hence, the gauge structure of systems with a local continuous symmetry can be encoded in a Matrix Product Operator (MPO) as depicted in Fig.~\ref{MPSplot}, such that:
\begin{eqnarray}
|\Psi_{QLM} \rangle & = &\sum_{\vec \alpha} A_{\alpha_1}^{\beta_1} A_{\alpha_2}^{\beta_1,\beta_2} \dots A_{\alpha_N}^{\beta_{N-1}} \\
 &&B_{\alpha_1, \alpha_1'}^{\gamma_1} B_{\alpha_2,\alpha_2'}^{\gamma_1,\gamma_2} \dots B_{\alpha_N,\alpha_N'}^{\gamma_{N-1}} 
| \alpha_1 \alpha_2 \dots \alpha_N \rangle,\nonumber.
\label{LGTN}
\end{eqnarray}
For the model we study, the MPO has bond dimension three ($\gamma=1,2,3$) and the non-zero elements of the tensors $B$ for the odd sites can be expressed as:
\begin{equation}
\begin{split}
& B_{1,1}^{\gamma_\imath,\gamma_{\imath+1}} = \begin{pmatrix} 0 & 0 & 0 \\ 0 & 0 & 1 \\ 0 & 0 & 0 \end{pmatrix}  ~B_{2,2}^{\gamma_\imath,\gamma_{\imath+1}}  = \begin{pmatrix} 0 & 1 & 0 \\ 0 & 0 & 0 \\ 0 & 0 & 0 \end{pmatrix};  \\
&B_{3,3}^{\gamma_\imath,\gamma_{\imath+1}}  = \begin{pmatrix} 0 & 0 & 0 \\ 0 & 1 & 0 \\ 0 & 0 & 0 \end{pmatrix} ~ B_{4,4}^{\gamma_\imath,\gamma_{\imath+1}}  = \begin{pmatrix} 0 & 0 & 0 \\ 0 & 0 & 0 \\ 0 & 0 & 1 \end{pmatrix}; \\
&B_{5,5}^{\gamma_\imath,\gamma_{\imath+1}} = \begin{pmatrix} 1 & 0 & 0 \\ 0 & 0 & 0 \\ 0 & 0 & 0 \end{pmatrix}.
\label{MPO}
\end{split}
\end{equation}
The tensors for the even sites can be computed in a similar way. Given the gauge invariant tensor structure introduced above, one can reformulate the standard algorithms for tensor networks to compute the ground state of the model, or, as we will do in the next Sections, compute the real time evolution of some initial state, e.g., by means of a Suzuki-Trotter decomposition of the time evolution operator acting on a pair of neighboring sites~\cite{Vidal,Daley}. The parameters used in our calculation, together with a discussion of the errors involved in the approximations, are presented in Appendix~\ref{App}.
Finally, we mention that one can further simplify the tensor structure in Eq.~\eqref{LGTN} to increase the algorithmic efficiency and that this construction is completely transparent with respect to the Abelian or non-Abelian nature of the gauge symmetry, resulting in a drastic simplification of numerical analysis of non-Abelian lattice gauge theories~\cite{Rico,Silvi2}. 

\section{\label{sec:StrBreaking}String Breaking}

In this section we present our results of the lattice gauge TN numerical simulations for the real-time dynamics of string breaking: we focus on the time evolution of the electric and matter fields, quantitatively studying the properties of string breaking and of the Schwinger mechanism. In the following section we analyze the time evolution of quantum correlations during this process -- an analysis enabled by the TN approach -- and show that the two figures of merit are intimately related. 

The setup for our simulations is a dynamical string surrounded by the vacuum. The total lattice length $N=100$ is chosen such as boundary effects are negligible for the timescales we investigate. Some typical results of the electric field time evolution are shown in the left column of Fig.~\ref{Overview:fig} (Column A) for different values of the fermion mass $m$ and electric field coupling strength $g$ (hereafter we set $t=\hbar=1$ and times are given in units of $\hbar/t$). In the top row, the string freely breaks as for $g=m=0$ no energy-cost is needed for such process to occur and the system evolves according to the free hopping Hamiltonian: the electric field starts to oscillate between string and anti-string displaying primary, secondary (marked by vertical lines) and subsequent string breakings. Moreover, the string propagates into the vacuum creating two diverging electric field excitation wavefronts. 
In the middle row, representing the same scenario for non-zero mass and electric field coupling strengths, string breaking still occurs; however, the string evolution is damped and stabilizes the mean electric flux around zero. Finally, for large mass $m$ and $g$, the large mass suppresses vacuum fluctuations while the large electric coupling prevents the occurrence of string breaking and thus the string does not decay. However, some dynamics still occur as we will see in more details in subsection~\ref{sec:Schwing}, as fermion-antifermion pairs are created using the energy of the electric field and then annihilated, thus restoring the electric field excitations.

\begin{figure}[t]
\begin{picture}(200,230)
\put(-15,0){\epsfig{file=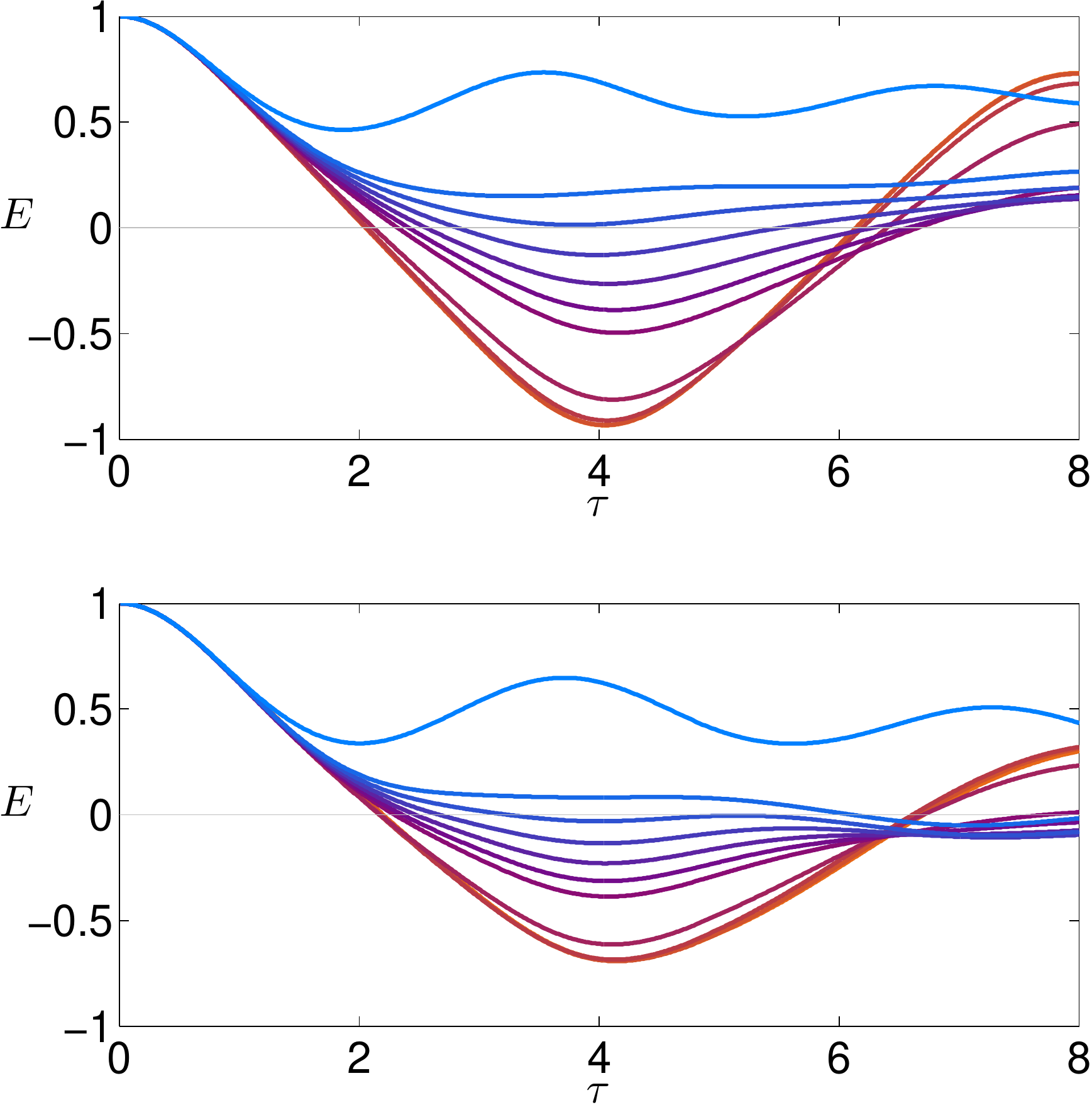,width=8cm,angle=0,clip=}}
\end{picture}
\caption{Mean electric field of the central six lattice sites as a function of time $\tau$ for the electric coupling $g=0.00, 0.25, 0.50, 0.75, 1.00, 1.05, 1.10, 1.15, 1.20, 1.25, 1.50$ (orange to light blue) for $m=0$ (top) and $m=0.25$ (bottom). Primary (secondary) string breaking occurs when the mean electric field crosses the zero-line from positive (negative) values.}
\label{StringBreakingEvolution:fig}
\end{figure}

\begin{figure}[t]
\epsfig{file=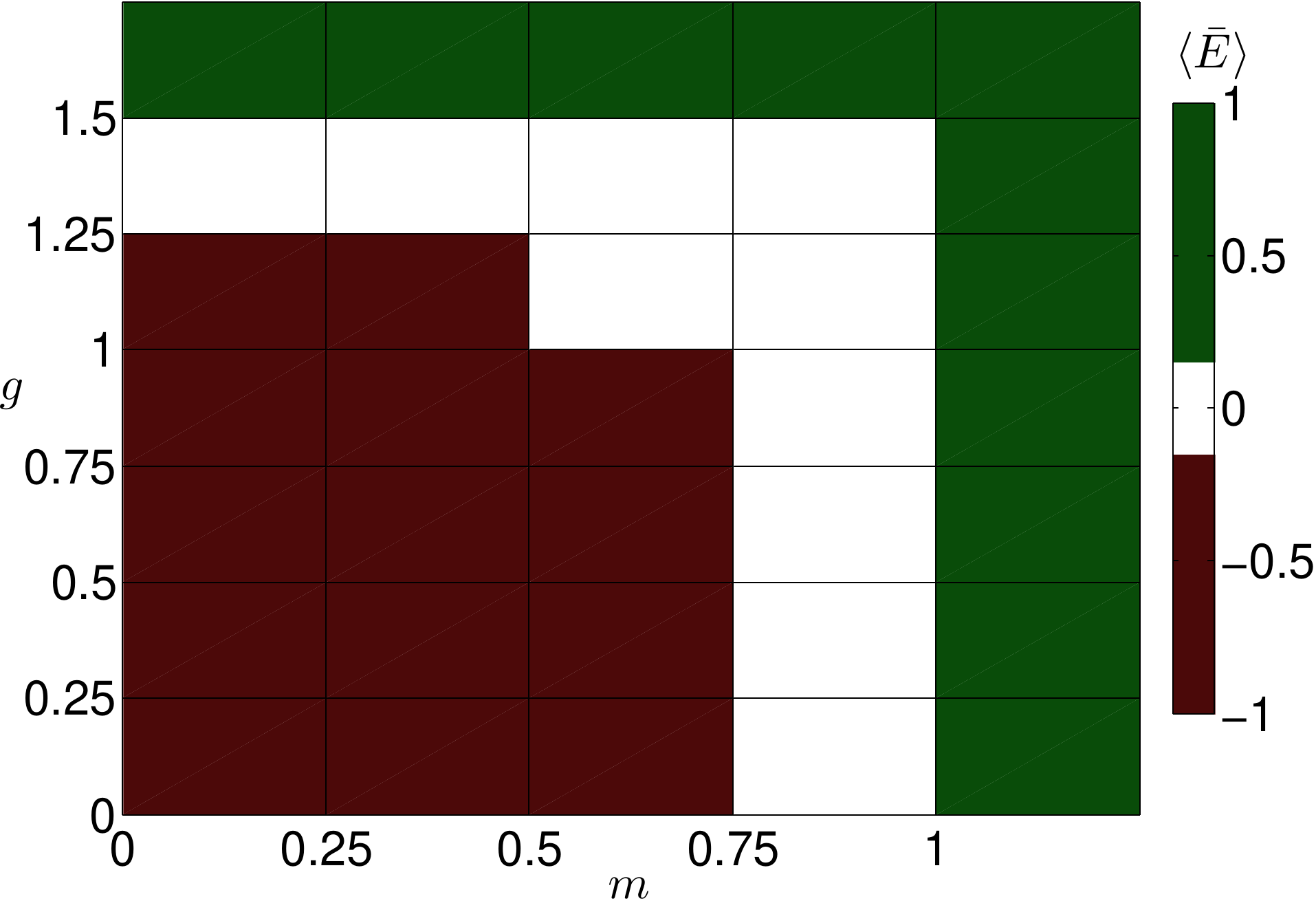,width=7cm,angle=0,clip=}
\caption{State diagram of string breaking. The red area shows the parameters where the string breaks and evolves into a negative string ($E_{\mathrm{mean}}<-0.15 E_{\mathrm{max}}$). The white area represents the parameters where the mean electric field approaches zero and stays around that value ($|E_{\mathrm{mean}}|<0.15 E_{\mathrm{max}}$). And finally, the green area represents the parameters without string breaking ($E_{\mathrm{mean}}>0.15 E_{\mathrm{max}}$).}
\label{StringBreakingMap:fig}
\end{figure}

To perform a quantitative analysis of string breaking, we repeat the same simulation for different values of $m$ and $g$ and analyze the mean electric field of the central six lattice sites of the string as a function of time. The results are reported in Fig. \ref{StringBreakingEvolution:fig} where one can clearly see that different scenarios might occur: either the string breaks when the mean electric field drops below zero, or the electric flux remains positive throughout the whole evolution and no string breaking occurs. In the limit of $m=g=0$ the oscillations show the highest amplitude, which is reduced by changing at least one of the two parameters; as previously noticed, for high values of either system parameter the string never breaks. In the regime between these two extreme cases, we observe a third type of behavior: the electric flux tends to zero, however no anti-string is formed: the oscillation is strongly damped and the system remains in the broken string state (compare with the middle row in Fig. \ref{Overview:fig}).
These findings are summarized in Fig. \ref{StringBreakingMap:fig}: For $g,m\lesssim 1$ we observe the full string breaking dynamics with at least the partial formation of a string with opposite electric field flux after reaching the broken string state (red area). For $g,m\gtrsim 1$ string breaking was not observed and the dynamics are dominated by the interplay of the state of maximum pair creation and the original string (green area). Finally, the white area in between represents the region of parameters where we observe the string breaking with over-damped oscillations.

\subsection{\label{sec:Spread}String wavefront spreading}
During the string breaking process, a wavefront of electric flux spreads outwards, as can be clearly seen in Fig. \ref{Overview:fig} (Panel A1 and A2). In this Section, we quantitatively characterize the wavefront spreading by analyzing its spreading velocity and the oscillation intensity.

\begin{figure}[t]
\epsfig{file=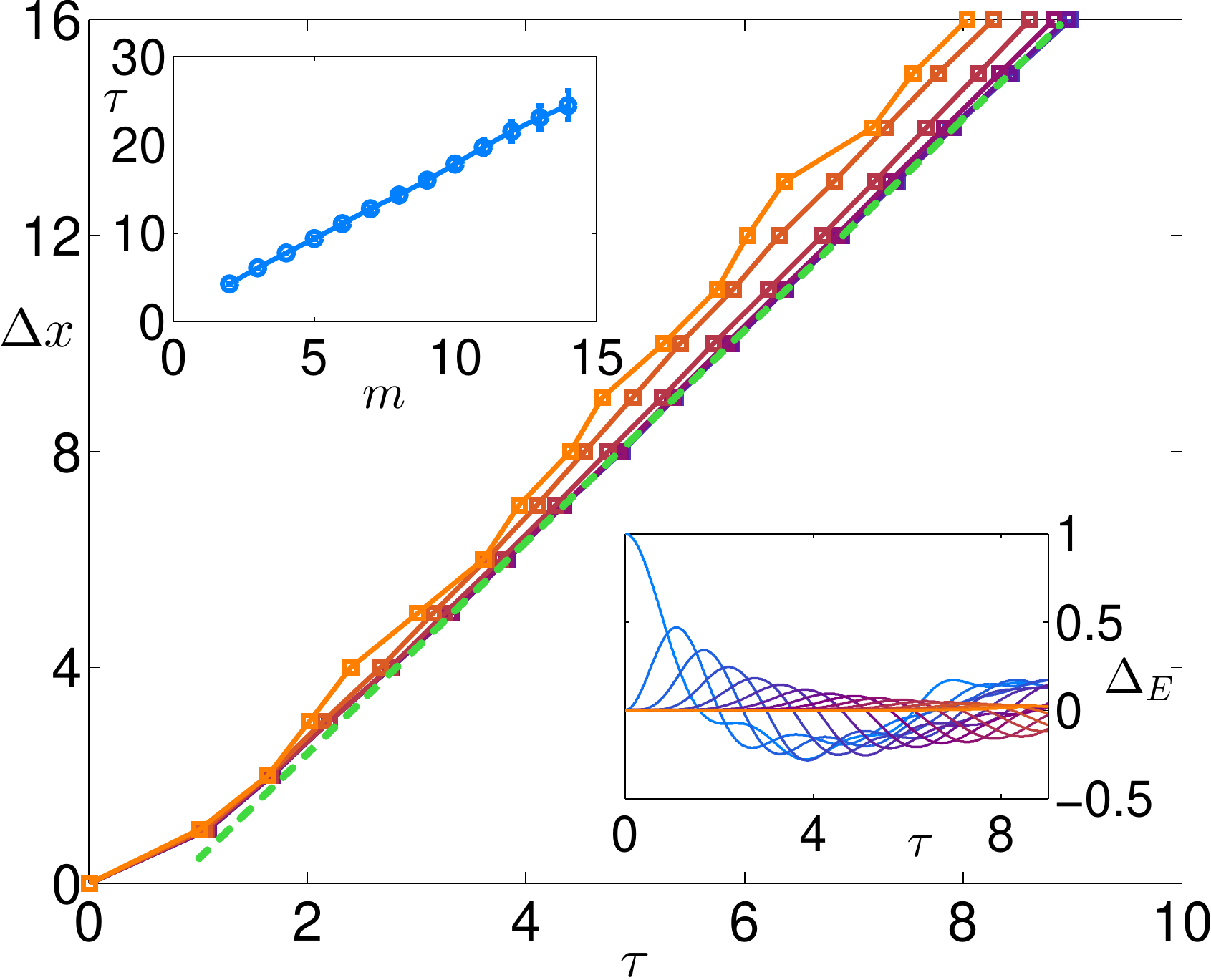,width=8cm,angle=0,clip=}
\caption{Spreading of the wavefront of the electric field for $m=0$ and $g$ increasing from $g=0$ (blue) to $g=1.5$ (orange). A linear fit for $\tau>2$ results in a velocity of $v_{\mathrm{E}}=1.96 \pm 0.02$ (dashed green line). Inset (top left): Time needed for the wavefront to spread one lattice site as a function of $m$ for $g=0$. Inset (bottom right): Wavefront as a function of time evolving from the last site of the original string ($x=x_i$, blue) to the lattice site $x=x_i+16$ (orange) for $g=0$. The wavefront is calculated using the electric field difference $\Delta_E=E(x+2)-E(x)$.}
\label{SpreadingEF:fig}
\end{figure}

In Fig. \ref{SpreadingEF:fig}, we show the wavefront propagation as a function of time for different electric coupling $g$ for the zero mass case. The lower inset illustrates how we calculated such propagation: we follow the electric field excitation on one side of string by means of tracking the difference $\Delta E$ between the gauge field at some position $x$ and the next nearest neighbor site $x+2$. Further, we define the time when this difference displays a maximum s arrival of the wave front ~\footnote{In order to avoid even-odd effects due to the staggered nature of the fermion fields, we have used odd sites here for the sake of clarity}. As can be seen from the lower inset of Fig. \ref{SpreadingEF:fig}, where different colors represent different coordinates $x$, a wave-like propagation can be clearly identified.

Following this scheme, we plot the position of the wavefront as a function of time in the main panel of Fig.~\ref{SpreadingEF:fig}. The result shows an approximatively linear spreading after an initial transient time of about $\tau\approx 2$, with a velocity almost independent of the values of the electric coupling for $g<1$. Increasing $g$ starting from $g=1$, the velocity increases as well, until for $g>1.5$, where the results are inconclusive as increasing $g$ leads to smaller wavefront amplitudes and consequently the errors bars prevent an accurate analysis to be carried out. 
However, for small $g$, the spreading velocity can be extracted directly from Fig.~\ref{SpreadingEF:fig}. By fitting the values for $\tau>2/t$ and $m=0$ we obtain a value of $v_{\mathrm{E}}=1.96 \pm 0.02$. 

Finally, in the upper inset of Fig. \ref{SpreadingEF:fig}, we repeated this analysis for different masses and $g=0$. The results clearly show that, for sufficiently large $m/t\gtrsim 4$, the wave front spread velocity has an inverse linear dependence on the mass.
All these results are in agreement with a theoretical estimate obtained assuming the ends of the string as sources of excitations: in a quasi-free or weak coupled model, the speed is related with the band-width of the kinetic term resulting on an excitation spreading velocity proportional to $v_{th} =2/m$.

\begin{figure}[t]
\epsfig{file=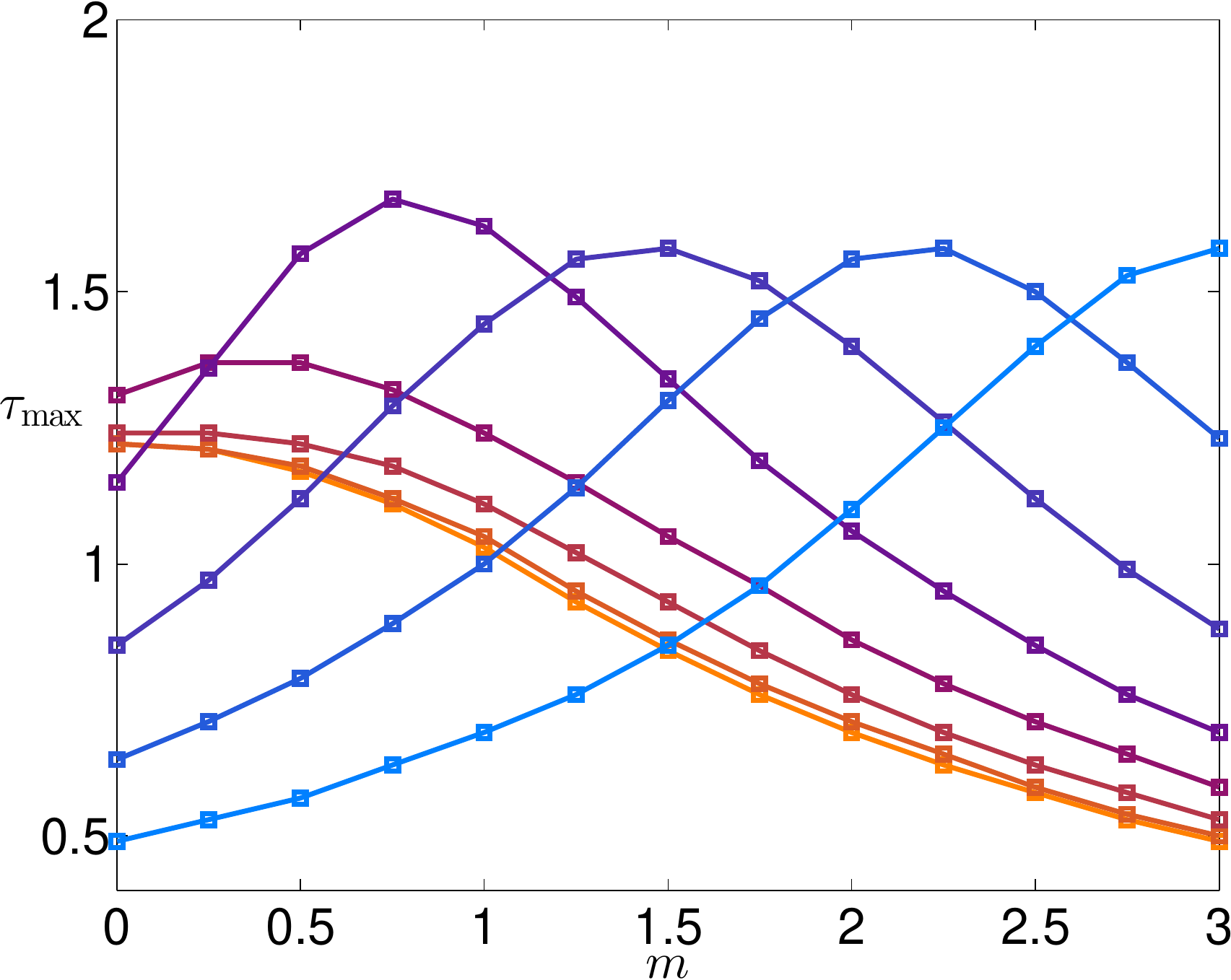,width=8cm,angle=0,clip=}
\caption{Time needed for the maximal mass production caused by the Schwinger mechanism as a function of the mass for $g =0,0.5,1,1.5,2,2.5,3,3.5$ (orange to light blue).}
\label{SchwingerTimes:fig}
\end{figure}

\subsection{\label{sec:Schwing}Schwinger mechanism}

During string breaking, the Schwinger model dynamics exhibit particle-antiparticle pair production as a consequence of the energy released from the external electric field string. This phenomenon is usually referred to as the Schwinger mechanism, and has been studied extensively since its first presentation in 1951~\cite{Schwinger0}. In the following, we provide a systematic investigation of the Schwinger mechanism in the context of the U(1) QLM.

\begin{figure}[t]
\epsfig{file=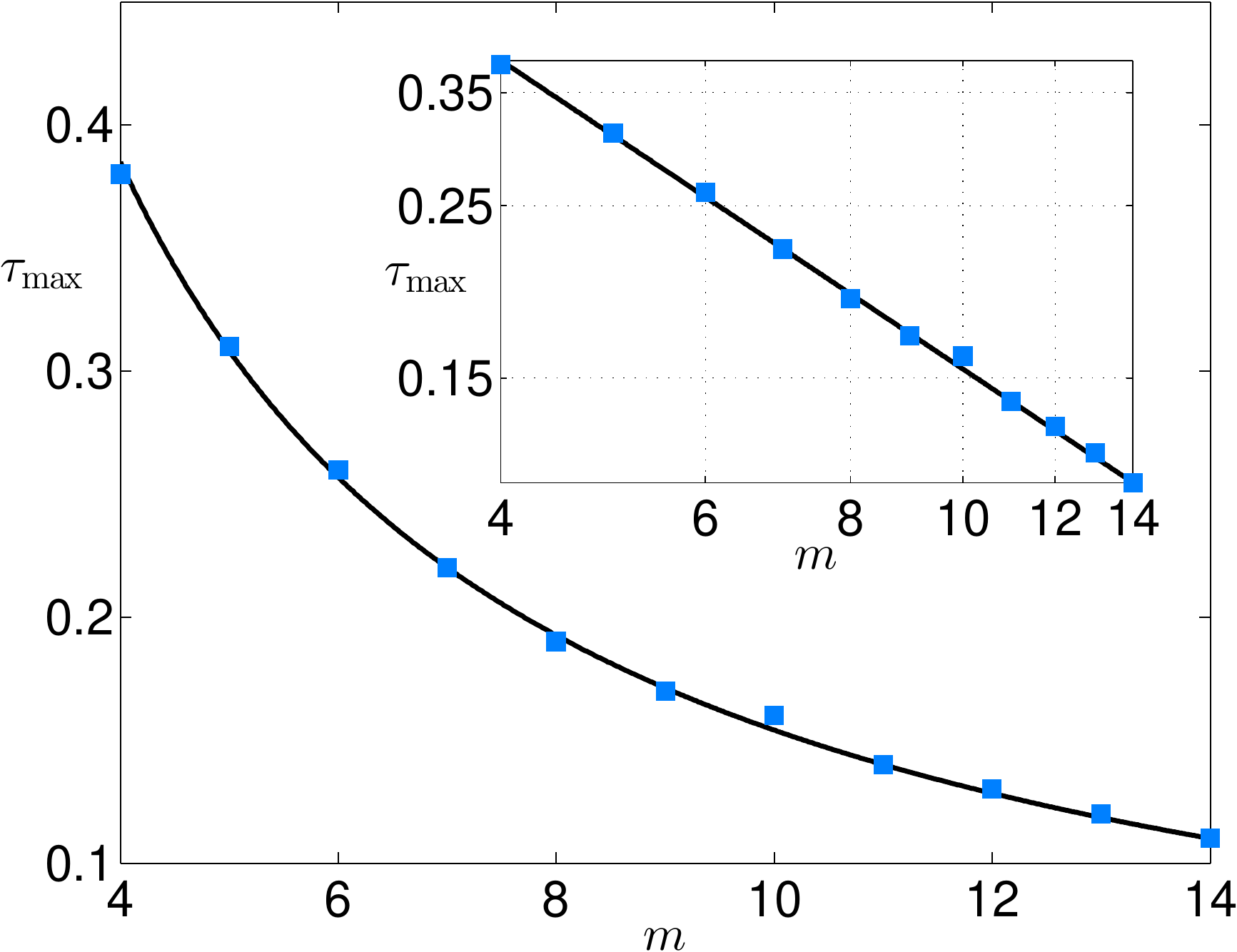,width=8cm,angle=0,clip=}
\caption{Large-mass behavior of the mass-production time-scale for $g=0$. The black line fits the data points with $\tau_{\mathrm{max}}=(1.54\pm 0.02)/m$. Inset: Log-log plot of the same data.}
\label{SchwingerTimesHigh:fig}
\end{figure}

In our analysis, in the initial state defining the string, the only two mass excitations present are the two dynamical charges which create the string itself. However, during the dynamics, the energy of the string is transformed into mass excitations. When the maximum mass production is reached, the particles start to annihilate, either to break the string or to restore it. The time needed to reach the maximum mass production $\tau_{\mathrm{max}}$ depends on the mass and electric coupling as shown in Fig. \ref{SchwingerTimes:fig}. It clearly displays two different behaviors, depending whether the electric coupling $g$ is greater or less than one. For small-$g$, $\tau_{\mathrm{max}}$ is maximal for $m=0$ and decreases monotonically with $m$. This occurs, recalling the results of Fig. \ref{StringBreakingMap:fig}, in the regime where string breaking is observable. Indeed, these are the cases comparable to the top row in Fig. \ref{Overview:fig}. On the contrary, for larger values of $g$, we observe the maximum to occur for $m>0$. In particular, the maximum is obtained at a point where the energy of the electric field matches approximately the energy needed to fully convert the string into particle pairs, i.e., $m=g^2/4$. This corresponds to the dynamics as displayed in the bottom row in Fig. \ref{Overview:fig}. In the regime of high masses we can use second-order perturbation theory to estimate the mass-dependence of the mass-production time-scale analytically, as the model can be approximated as decoupled double-wells potentials, resulting in $\tau_{\mathrm{max}} = \frac{\pi/2}{m}\approx \frac{1.57}{m}$. We checked this approximation comparing the analytical estimate with the numerical results in Fig. \ref{SchwingerTimesHigh:fig}: The best fit results in $\tau_{\mathrm{max}} = (1.54\pm 0.02)/m$, in good agreement with the theoretical prediction.  

\subsection{\label{sec:Imp}Observability of string breaking in synthetic platforms}

Recently, the implementation of Abelian quantum link models has been envisaged on different platforms, such as ultra cold atom gases in optical lattices~\cite{Banerjee,Zohar2,Stannigel}, trapped ions~\cite{Hauke}, and circuit QED architectures~\cite{Marcos1,Marcos2}. In this context, the possibility of investigating the real-time dynamics using TN methods provides an invaluable tool to benchmark experiment against theory in (1+1)d, and to address the role of possible imperfections in quantum simulators. 

Unavoidable imperfections which will be present in any implementation can be detrimental to the observation of both ground state physics properties and real-time dynamics. The former case has recently been investigated in the context of adiabatic state preparation, where it was shown how gauge-variant perturbations weakly affect the fidelity of the loading process~\cite{Kuhn}. Here, we focus instead on the effect of gauge-invariant imperfections on the string-breaking dynamics. Different from gauge-variant terms, the role of gauge-invariant imperfections cannot be systematically addressed in an experiment using, e.g., post-selection over the experimental data. 

Following the implementation schemes in Ref.~\cite{Banerjee,Hauke,Marcos1}, one of the most common form of gauge-invariant imperfections are nearest-neighbor interactions between matter and gauge fields:
\begin{eqnarray}
 H_I&=&\xi\sum_x n_x\left( S^z_{x-1,x} + S^z_{x,x+1} \right).
   \label{eq:Imp}
\end{eqnarray}
with $n_x=\psi_x^\dagger\psi_x$. This form of the imperfection is usually generated as a resonant term in perturbation theory to next-to-leading order with respect to $t$. While this implies $t\gg \xi$, for realistic implementations the difference in magnitudes between these two terms cannot be made arbitrary large: this will require small absolute energy scales, thus making other sources of more detrimental imperfections such as temperature (for the cold atom implementation) and disorder (in the circuit QED implementation) dominant. At a qualitative level, this interaction term can freeze the system into a configuration where the matter fields remain pinned. This is due to the effective attraction generated by the nearby electric field configuration. 

\begin{figure}[t]
\epsfig{file=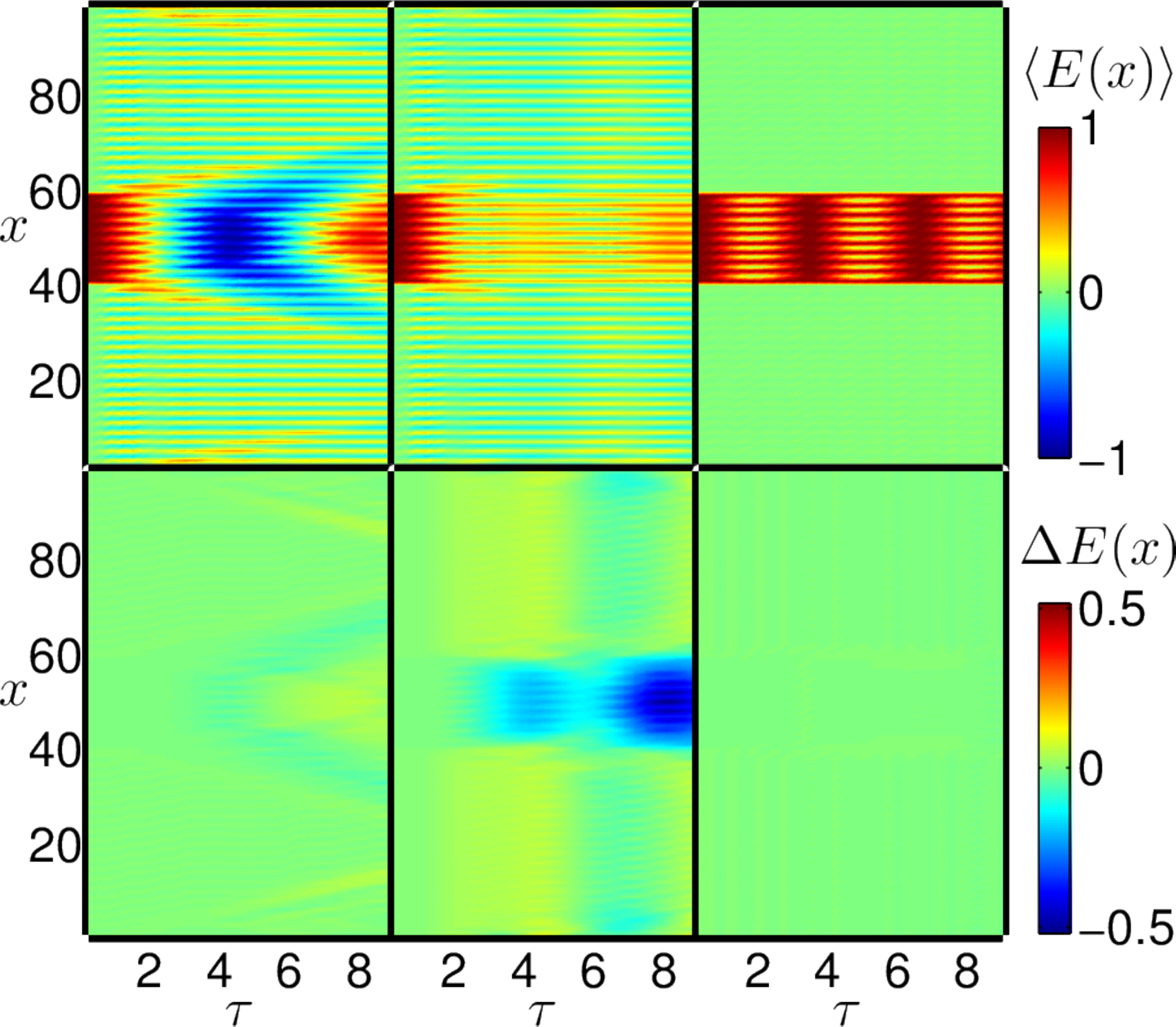,width=9cm,angle=0,clip=}
\caption{Time evolution of the electric field including imperfections of the type $H_I=\xi \sum_x n_x (S^z_{x-1,x}+S^z_{x,x+1})$ with $\xi=0.1$ at the system parameters $m=0$, $g=0$ (left), $m=0.25$, $g=1.25$ (center) and $m=3$, $g=3.5$ (right). The bottom row shows the difference to the result obtained without imperfection.}
\label{Imperf:fig}
\end{figure}

To estimate the effects of these imperfections on a quantum simulation of the dynamics considered in this work, we repeat the numerical simulations including realistic imperfections expected in a first generation of experiments. In the top row of Fig.~\ref{Imperf:fig} we show the string breaking evolution of the electric field, in the presence of $H_I$, with the same system parameters as used for Fig. \ref{Overview:fig}, with imperfections of the order of 10\% $\xi=0.1$. The results including the imperfections still clearly exhibit the physics observed in the imperfection-free results. In general, even the quantitative dynamics is very well captured up to long-timescales, as illustrated in the bottom row of Fig.~\ref{Imperf:fig}. The only exception are intermediate $g$ and $m$ values, where significant discrepancies (up to 50\%) are observed for intermediate timescales (middle panel of the last row). In all other regimes, we could observe deviations up to a maximum of 15\% caused by typical imperfections $\xi=0.1$.

\section{\label{sec:Ent}Entanglement dynamics}

One of the key aspects of MPS-based methods is that they give full access to the wave-function during the real-time dynamics.  By considering the dynamics of quantum correlations embodied by entanglement, this enables us to tackle the string-breaking problem from a fully complementary viewpoint with respect to the electric field and mass generation studies undertaken in the previous sections. The main question we want to address in this section is whether and to what extent entanglement plays a role in the string-breaking dynamics.

In the last decade it has been shown that entanglement play a fundamental role in many-body quantum processes, from quantum critical phenomena, to quantum information theory and  other fundamental aspects of quantum physics. Moreover, several aspects of quantum field theories, such as the static properties of conformal field theories, have also been extensively studied using entanglement measures~\cite{Holzhey,Calabrese}. Additionally, entanglement was shown to play a crucial role in the limits of classical simulations of quantum systems calling for the need of the development of quantum simulators to overcome such limitations~\cite{Schollwoeck1}.

A common way to quantify entanglement for pure quantum states is by using the so-called Renyi entanglement entropies, and, in particular, the von Neumann entropy. Given a pure state with the density matrix $\rho=| \psi\rangle\langle\psi |$, the entanglement entropy is given by the von Neumann entropy of the reduced density matrix $\rho(x)=\mathrm{Tr}_{L-x}\rho$~\cite{Calabrese}: 
\begin{equation}
S(x)=-\mathrm{Tr} \left\{ \rho(x)\log_2 \rho(x) \right\}.
\end{equation}
$S(x)$ is a measure of the entanglement of a bipartition at the lattice site $x$. The entanglement entropy takes values between $S(x)=0$ for a separable state (product state) and $S(x)=\log_2 d$, with $d$ being the size of the Hilbert space, for a maximally entangled state. 

\subsection{Von Neumann entropy after string breaking}

In Fig.~\ref{Overview:fig} (Column C) we plot the time evolution of the entanglement entropy for the three different cases considered before (Panels 1-3): as it can be clearly seen, the entanglement evolution resembles the mass and electric field excitations, indicating how the two phenomena are related. Firstly, the vacuum fluctuations for small $g$ and $m$ generate not only mass and electric field fluctuations, but also a high amount of correlations. Moreover, the electric field wavefront is mimicked by the entanglement behavior, once again showing that not only do excitations propagate from the string, but also correlations do as well. Secondly, the string breaking process is clearly a two-steps process: first a correlated pair is created in between odd-even sites, and later in between the even-odd sites (blue-yellow checkerboard pattern inside the string in Panels C1 and 2). Finally, for large $g$ and $m$, when the string does not break, the correlations behavior drastically changes as well. There, within the string, the correlations are built periodically only between odd-even sites, while in the vacuum the correlations are drastically suppressed (Panel C3).

\begin{figure}[t]
\epsfig{file=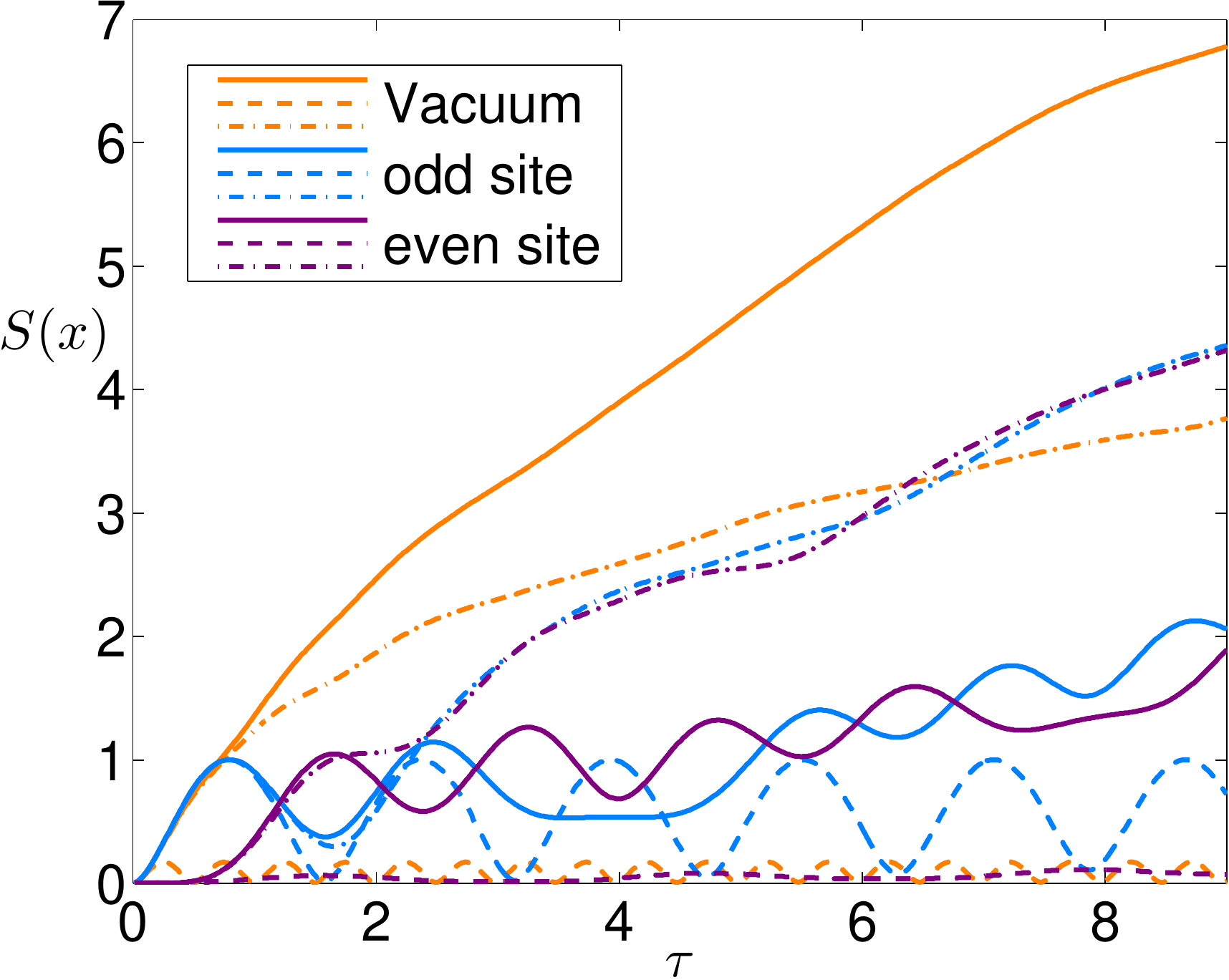,width=8cm,angle=0,clip=}
\caption{Bipartite von Neumann entropy $S(x)$ at $m=0$, $g=0$ (solid), $m=0.25$, $g=1.25$ (dot-dashed) and $m=3$, $g=3.5$ (dashed). Partition at the center of the initial string ($x=51 - $blue, $x=50 - $violet) and in the vacuum ($x=20 - $orange).
}
\label{EntropyProfile:fig}
\end{figure}

A transparent picture on how entanglement is generated during the quench dynamics can be gathered by monitoring the entanglement growth at different points in space. In Fig.~\ref{EntropyProfile:fig} we show the entanglement time evolution for a set of system bipartitions and parameters, e.g., cutting the system in the middle of the string or in the vacuum. 

The solid lines are results obtained for $m=0$ and $g=0$, while the dashed lines correspond to the case $m=3$ and $g=3.5$. The orange line represents the entanglement growth for a partition at lattice site $x=20$, therefore in the center of the lattice region starting in the vacuum. In the zero-mass case, one can see that the entanglement entropy grows almost linearly for most part of the evolution - signaling the vacuum instability against resonant particle-antiparticle pair production. Towards the end of the evolution, the linear growth breaks down as boundary effects start to play a role. The remaining solid lines in Fig.~\ref{EntropyProfile:fig} represent the behavior of a partition between an even-odd lattice site (violet) and between an odd-even lattice site (blue) and display a more complex behavior. These results were obtained in the center of the string while it breaks up: the counter-phase oscillations indicate the competition of two states, together with an overall growth of the entanglement entropy, though not as fast as in the vacuum. As we have seen, in this regime the string breaking is a result of consecutive hopping processes: fermions hop on the lattice creating mass excitations followed by annihilations, which result in the string breaking with two remaining dynamical mesons. This dynamical process goes on and after two hopping processes the anti-string is created. Fig.~\ref{EntropyProfile:fig} shows that these dynamics are well captured by the oscillations of the entanglement entropy. Each maximum represents one hopping process: at the first maximum of the blue line the fermions are about to hop for the first time, while at the first maximum of the violet line the second hopping process is at its peak resulting in the string broken state. At around $\tau \approx 4$, the blue line does not display a maximum: this 'depleted region' signals the anti-string state, where the last hopping event creating the anti-string is again the first hopping to break the negative string (and thus, it takes twice as long for the entropy to reach the next maximum).

\begin{figure}[t]
\epsfig{file=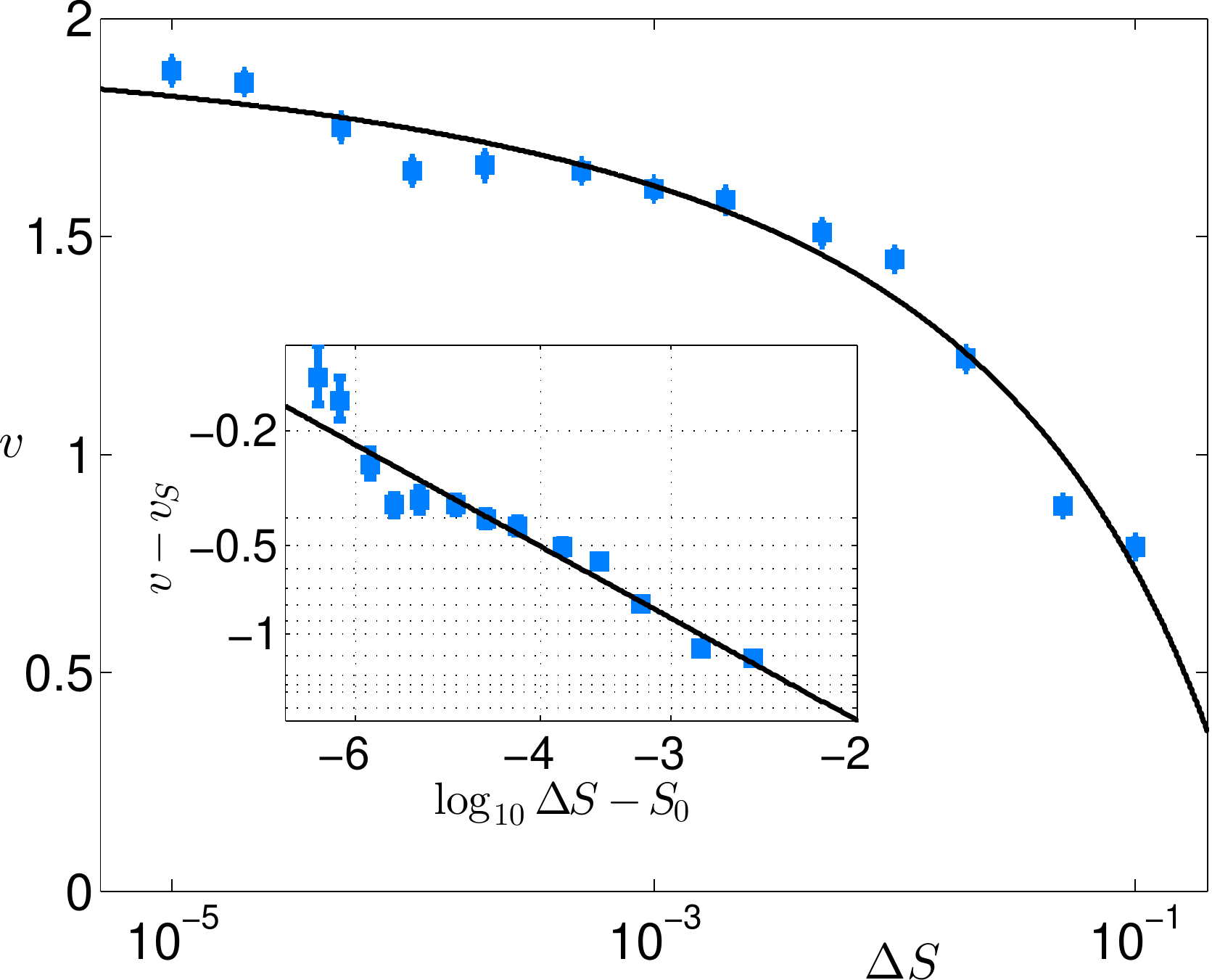,width=8cm,angle=0,clip=}
\caption{Estimation of the spreading velocity $v$ from the entanglement entropy $S$. The wavefront is defined as when the entropy drops below a certain threshold value with respect to the vacuum. The resulting fit using $v=v_{\mathrm{S}}-a/[\log_{10}(\Delta S)-S_0]^2$ gives us an estimate for the spreading velocity of $v_{\mathrm{S}}=(2.0\pm 0.2)$.
Inset: Log-log plot of the same data adjusted by $v_\mathrm{S}$ and $S_0$.}
\label{SpreadingEE:fig}
\end{figure}

Differently from before, in the massive scenario (dashed lines), we see that the entanglement entropy for the vacuum stays close to zero as the large mass and electric coupling strongly suppress the particle-pair creation that triggered the strong growth of the entropy in the previous case. Moreover, in the middle of the string the entanglement entropy is drastically affected: the blue dashed line initially behaves as the solid line in the massless case, reflecting the same mass excitation by pair creation. However, the violet dashed line always remains close to zero as further evolution into the string broken state is energetically forbidden: the state evolves back into the string and the correlations between the even-odd sites cannot be created. The system is then oscillating between two almost degenerate states: the initial string state and the state made out of pairs. This results in the oscillating behavior of the entanglement entropy between zero and one. Finally, the third case with $m=0.25$ and $g=1.25$ (dot-dashed lines) lies between the two previous limiting cases: here the string breaks, but does not evolve into an anti-string. In the vacuum, the entanglement evolution is very similar to the first case ($m=0=g$) as the entropy grows almost linearly after a transient regime. However, the slope is reduced by the nonzero mass. The entanglement in the center of the string initially evolves as for the massless case, but after the first two hopping processes for $\tau\gtrsim 2$, the oscillation turns into vacuum-like growth. This is a strong indication for non-periodic string breaking: the dynamics, although being unitary, resemble a dissipative process where the electric field energy irreversibly disperses into the vacuum. This irreversible behavior directly resembles what we observe in the electric field dynamics, which does not display any clear periodic signature.

\begin{figure}[t]
\epsfig{file=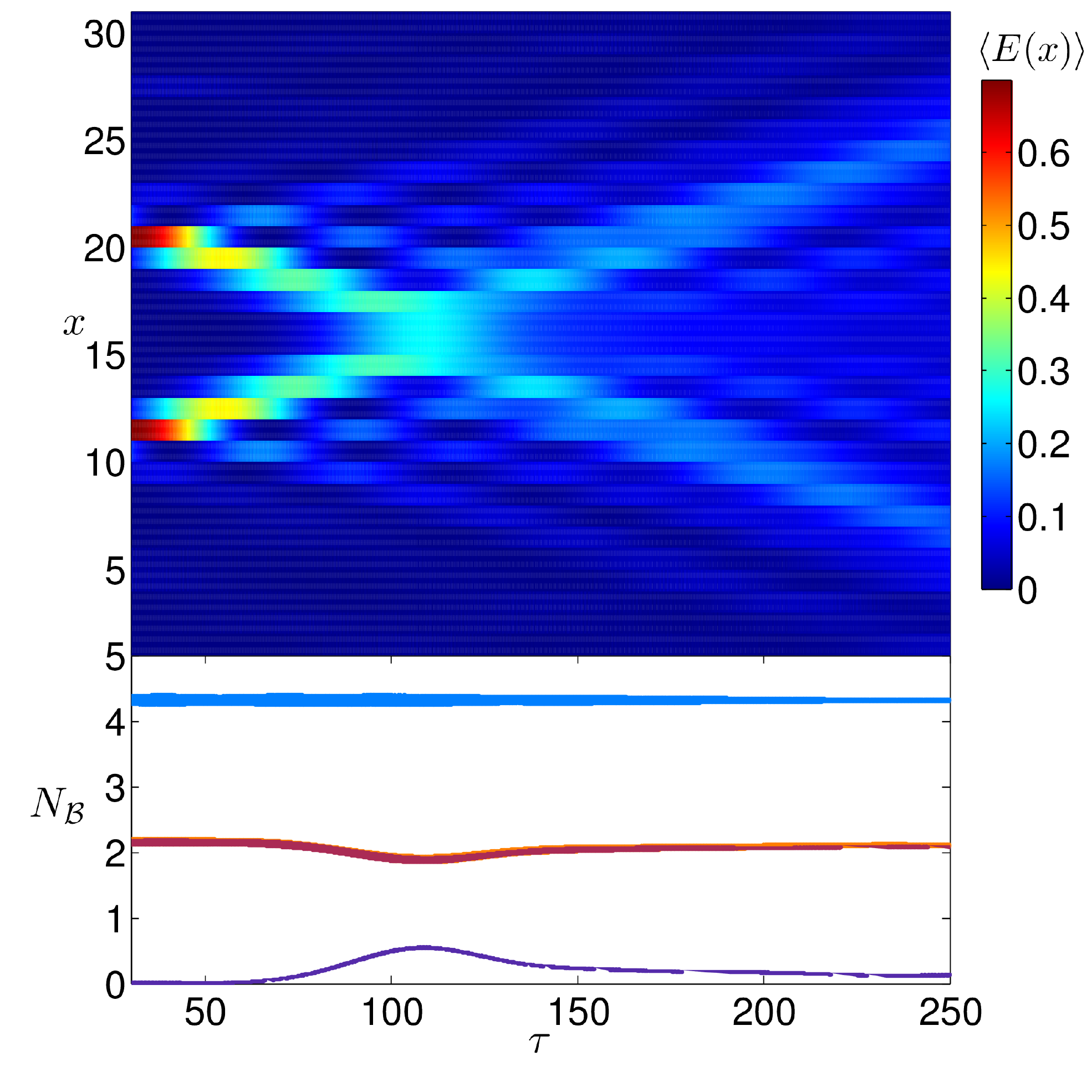,width=8.5cm,angle=0,clip=}
\caption{Scattering of two dynamical mesons using the system parameters $m=0$, $g=8$. The plot illustrates the time evolution of the electric field $E(x)$ as a function of the position $x$. Lower panel:
number of charges $N_{\mathcal{B}}=\sum_{x \in \mathcal{B}} n_x$ in the system during the evolution (blue: $\mathcal{B}=\{1, \dots ,32\}$), number of particles present in the center (purple: $\mathcal{B}={16}$), number of charges on either side of the center (coinciding lines red: $\mathcal{B}=\{1,\dots ,15\}$ and orange: $\mathcal{B}=\{17,\dots ,32\}$). }
\label{Scatt:fig}
\end{figure}

\subsection{Entanglement propagation and wavefront}

Even more remarkably, the real-space particle creation and the entanglement dynamics are quantitatively tied. We concentrate on the signatures of the wavefront of the string imprinted on the evolution of the entanglement entropy. We consider the case $m=g=0$ as it is characterized by the most pronounced wavefront, where the string with its slow entanglement growth is embedded in the fast growing vacuum (see Fig. \ref{Overview:fig}, panel C1). To characterize the entanglement spreading due to the wavefront, we exploit the fact that the entanglement entropy in the vacuum is constant in space even though it evolves in time. Therefore, far enough from both sides of the string there is a plateau of constant entropy much higher than the entropy in the middle of the string. Thus, to define the wavefront of entanglement spreading due to the string, one can look for the lattice site at which the entropy plateau starts to decrease. We identify this point computing the difference of entropy between nearest neighbor bipartitions: tracking when this quantity become bigger than a given threshold allows us to characterize the entanglement wavefront spreading. 

In Fig. \ref{SpreadingEE:fig} we show the estimated spreading velocity for different values of the threshold: the limit for the threshold value going to zero gives an estimate of the spreading velocity. A power law fit results in a spreading velocity of $v_S=2.0\pm0.2$ in very good agreement with the analytic estimate of $v_T\simeq 2$ and the result from the electric field of $v_E=1.96 \pm0.02$ demonstrating the intimate connection between entanglement and electric field spreading.

\begin{figure}[t]
\epsfig{file=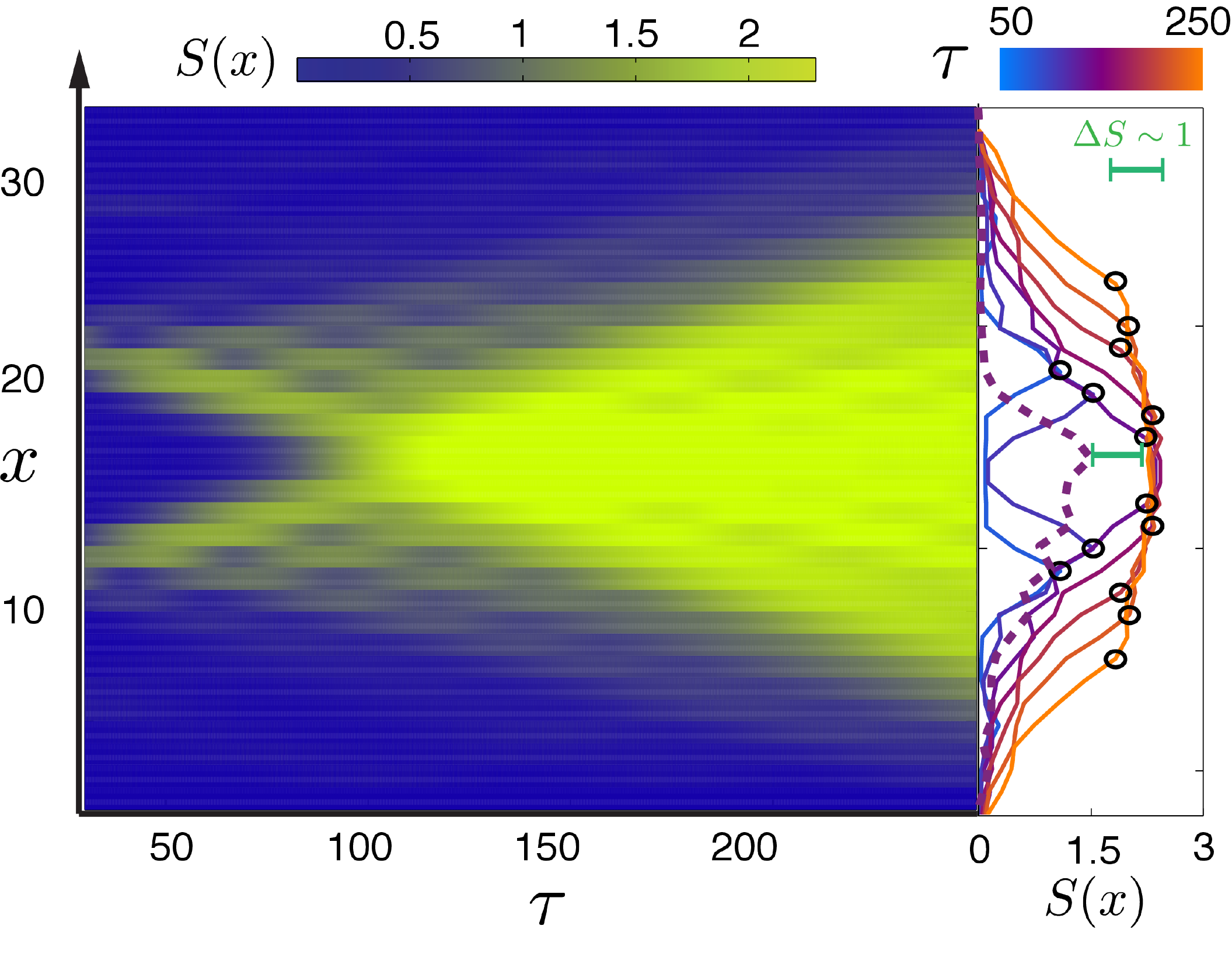,width=8cm,angle=0,clip=}
\caption{Scattering of two dynamical mesons. Main panel: Entanglement entropy $S(x)$ using a bipartition between sites $x$ and $x+1$ as a function to time. After the scattering, the entropy significantly increases in the system: this is a direct signature of enhanced quantum correlations. Right panel: $S(x)$ at different times (see color bar), showing a clear plateau after the collision, which enlarges as a function of time. The empty circles show the current position of the maxima of the electric-field which follow approximately the mesons center of mass. The dashed line represents $S(x)$ generated by a single meson, while the green bar highlights the difference $\Delta S$ to the entropy of the colliding mesons (difference between full and dashed line at $\tau=120, x_i=17$).}
\label{Escheat:fig}
\end{figure}

\section{\label{sec:Scatt}Meson scattering and entanglement generation}

Bound states are a fundamental component of gauge theories, and understanding their complex internal structure is one of the most ambitious goals of computational physics~\cite{fodor2008}. In experiments, such internal structure is usually explored by colliding heavy ions, so that the energy released during the process can be released via particle-antiparticle creation. This makes the {\it ab initio} numerical simulations of such scattering processes challenging, as MC simulations suffer from a severe sign problem when tackling real-time dynamics. 

Here, we show how TN simulations allow to investigate meson scattering for the (1+1)d QLM using TNs. First, we present a general procedure to implement scattering processes between composite particles, and discuss the electric field dynamics after the collision. Then, we present results for the entanglement dynamics during and after the collisions showing that the meson collision is accompanied by the creation of entanglement between the mesons themselves. Indeed, as we will show, the entanglement is bounded by the propagation wavefronts of the particles after collision, and is characterized by a constant plateau of the entanglement entropy within the region.

\subsection{Electric field patterns during meson collisions}

In order to produce the scattering process, we shall start with two composite particles. Each particle is charge/anti-charge pair, and divided only by one link, namely a meson, with opposite momentum. For the two-meson problem, there is a simple picture in the strong coupling limit: the massless theory is a free massive boson (meson) theory that is expected to become weakly interacting once a small mass term is included. Hence, in the strong coupling region, a possible two-meson bound state is loosely bound, while in the weak coupling region it is tightly bound.

We start the numerical simulation with the state represented in the cartoon (D) in Fig.~\ref{summary:fig}: two mesons separated by a vacuum state of ten sites,  which can be straightforwardly be written in a simple, separable matrix product state with $t=0$. 
We provide momentum to the mesons by adiabatically moving them from the boundaries toward the center of the system: this is done by introducing a deep box-shaped potential that decouples the mesons from the rest of the system, with the only dynamics allowed being the oscillate between its position and a neighboring site. The box-potential is removed at time $\tau_i=17.4$ when the meson is exactly at half oscillation: from that point on the mesons evolve freely with an effective momentum mostly in one direction, one towards the other and eventually colliding~\footnote{As can be seen from Fig. \ref{Scatt:fig}, the meson wave packets still spread slightly in both directions: however the centers of mass of the two mesons are clearly propagating with constant and opposite momentum until the scattering process occurs and the particles bounce back. }. In order to avoid vacuum fluctuations during the process, we choose a large value of $g=8$.
Fig.~\ref{Scatt:fig} shows an example of such a scattering process. In particular, it shows the absolute value of the electric field of two mesons approaching each other, colliding in the center and there parting again. While before the collision the meson are tightly bound, after the scattering process the electric field diffuses, and the corresponding wavefront has a significantly attenuated signal. In the lower panel of Fig.~\ref{Scatt:fig}, we monitor the time-evolution of the total particle number (blue), clearly indicating that this quantity is approximately conserved over the entire time-evolution, due to the large electric field strength, which suppresses particle-antiparticle creation.

\subsection{Post-collision entanglement generation}

A classical-like picture of the scattering process presented above, would have that two particles move against each other and then bounce back as there is not enough energy available to generate a more complex inelastic scattering. However, this picture is oversimplified, as this is a fully quantum process. Indeed one can, once more, monitor the quantum correlations generated during the scattering process. This is done in Fig.~\ref{Escheat:fig}, where we show the evolution of the bipartite entanglement entropy: one sees that entanglement is created and that it is mostly carried by the two mesons. In this parameter regime, the vacuum does not generate entanglement due to the very large value of $g^2$. Studying the bipartite entanglement entropy for different bipartitions and times, one clearly sees that there are two regimes: before the scattering occurs, the entanglement is present only in the bipartition that cuts the mesons wave packets, indicating two  electron-positron wave packets internally correlated but not sharing any quantum correlations among them. To the contrary, after the scattering, the two wave packets become highly correlated even when their two centers of mass are clearly separated (see Fig.~\ref{Scatt:fig} for times $\tau > 100$).

The values of the entanglement entropy indicate that one ebit of quantum information has been created during the scattering process. 
In the right panel of Fig.~\ref{Escheat:fig}, we present various cuts of the entanglement entropy profile taken at different times, together with a comparison with the entanglement generated by a single meson moving through the lattice (dashed line). 
The difference of $\Delta S \approx 1$ between the two cases (highlighted in Fig. \ref{Escheat:fig}, green bar) clearly shows that one additional {\it ebit} of entanglement has been generated during the scattering process: that is, that a singlet state has been created between the two indistinguishable mesons. The entropy has increased as the information on which process occurs (either mesons bouncing and moving back, or each one moving freely through the other one and keeping moving with its own momentum) is completely lost. Notice that this is a direct consequence of the wave nature of the mesons: the classical case of two scattering particles in one dimension - with the constraint that no double occupancy might occur in a single matter site - would necessary results only in backscattering, since the two hard classical particles cannot pass through each other.

\section{\label{Sec:Concl}Conclusions}

We presented a detailed tensor network study on the real-time dynamics of a lattice gauge theory in the presence of dynamical charges and quantum gauge fields. Within this approach, we have shown that one has direct access to all local quantities of interests - the time evolution of the mass, charge and gauge fields - and to the quantum correlation between bipartitions of the system by means of the von Neumann entropy. We investigated the primary and secondary string breaking in QED in (1+1)d represented by an $S=1$ quantum link model with staggered fermions. In this context, we studied the real-time evolution of the Schwinger mechanism leading to mass creation and annihilation by means of the interplay with the electric energy released by the string. We quantified key properties of these effects such as the mass production rate of the Schwinger mechanism and the velocity of the electric field spreading. Moreover, we unveiled the relation between string breaking dynamics and the entanglement spreading in the systems and we showed that it is possible to study scattering dynamics, characterizing not only mass and charge real-time evolution but also the creation of quantum correlations among scattered particles. 
Finally, we showed that the presented results can be in principle verified experimentally in possible future quantum simulations, as they appear to be robust with respect to the most common sources of gauge-invariant imperfections appearing in most of the proposed implementations. 

This work paves the way to systematic studies of real time dynamical phenomena in Abelian and non-Abelian LGTs in low dimensional systems. Indeed, the present approach can be straightforwardly generalized to more complex LGT and geometries, e.g., ladders or cylinders, and also can be studied in presence of an external environment by means of, for example, the tensor network approach presented in~\cite{MPDO}. Moreover, one can also study the continuum limit of the LGT (as already discussed for Wilson theories~\cite{Banuls2}): for QLMs, this can either be done using dimensional reduction~\cite{WieseReview}, or by increasing the quantum link representation, similarly to what has been done in~\cite{Banuls1}. Notice that the gauge invariant tensor network formulation behaves favorably as the speed-up it grants
scales as the number of rishons per link squared~\cite{Silvi2}.
The unprecedented access to the entanglement dynamics in LGTs will allow investigations on the role of quantum correlations in different contexts enabling a deeper understanding of the quantum real-time dynamics of lattice gauge theories. 
Finally, exploiting the capability to prepare a wide class of complex states granted from recent developments in quantum optimal control of many-body quantum systems~\cite{OCT,OCT2}, more complex dynamics could be investigated. One example would be to perform extensive studies of the scattering at different energies. 

Alongside, these methods can also be applied to study condensed matter systems to compute, e.g., response
functions of antiferromagnets described by LGT (spin ices, Resonating Valence Bond models, etc.) which are very hard to evaluate using MC due to analytic
continuation~\cite{Lacroix2010}. Moreover, real time dynamics of gauge theories is fundamentally interesting to ab initio investigations of scattering equilibration and pre-thermalization~\cite{WieseReview}. 
Finally to benchmark and verify small quantum simulations whose proposal have recently appeared for different
platforms (ranging from cold atoms to superconducting circuits and trapped ions) and that can be foreseen to be experimentally
implemented in the next years~\cite{Weimer,Kapit,Zohar0,Zohar1,Banerjee,Zohar2,Marcos1,Hauke,Tagliacozzo2,Zohar3,Marcos2,Stannigel,Glaetzle,Notarnicola,Meurice}.

While writing this manuscript, we became aware of a related work on string breaking using TN methods~\cite{banuls2015}.

\subsection*{Acknowledgements}

We acknowledge useful discussions with D. Banerjee, J. Berges, P. Hauke, F. Hebenstreit, V. Kasper, E. Martinez, T. Monz, S. Pascazio, and U.-J. Wiese, and thank C. Laflamme and M. Rider for proofreading the manuscript. MD,  SM and PZ would like to acknowledge the hospitality of the Institute of Nuclear Theory - University of Washington, where this work was finalized.

Work in Innsbruck is partially supported by ERC Synergy Grant UQUAM, SIQS, and the SFB FoQuS (FWF Project No. F4016-N23), in Ulm by the EU projects SIQS and RYSQ, and by the DFG project SFB/TRR21, while in Bilbao, we acknowledge financial support from Basque Government Grants IT472-10 and IT559-10, Spanish MINECO FIS2012-36673-C03-02, UPV/EHU UFI 11/55, PROMISCE and SCALEQIT European projects.

\appendix
\section{Convergence checks of the MPS simulations}
\label{App}
We investigated String-breaking in a one-dimensional system with up to 100 lattice sites. The simulation was performed using a matrix-product-state (MPS) algorithm using a second order Suzuki-Trotter decomposition for the time-evolution. In this appendix we present the parameters we used for our simulations and provide a discussion on the relative errors.

\begin{figure}
\epsfig{file=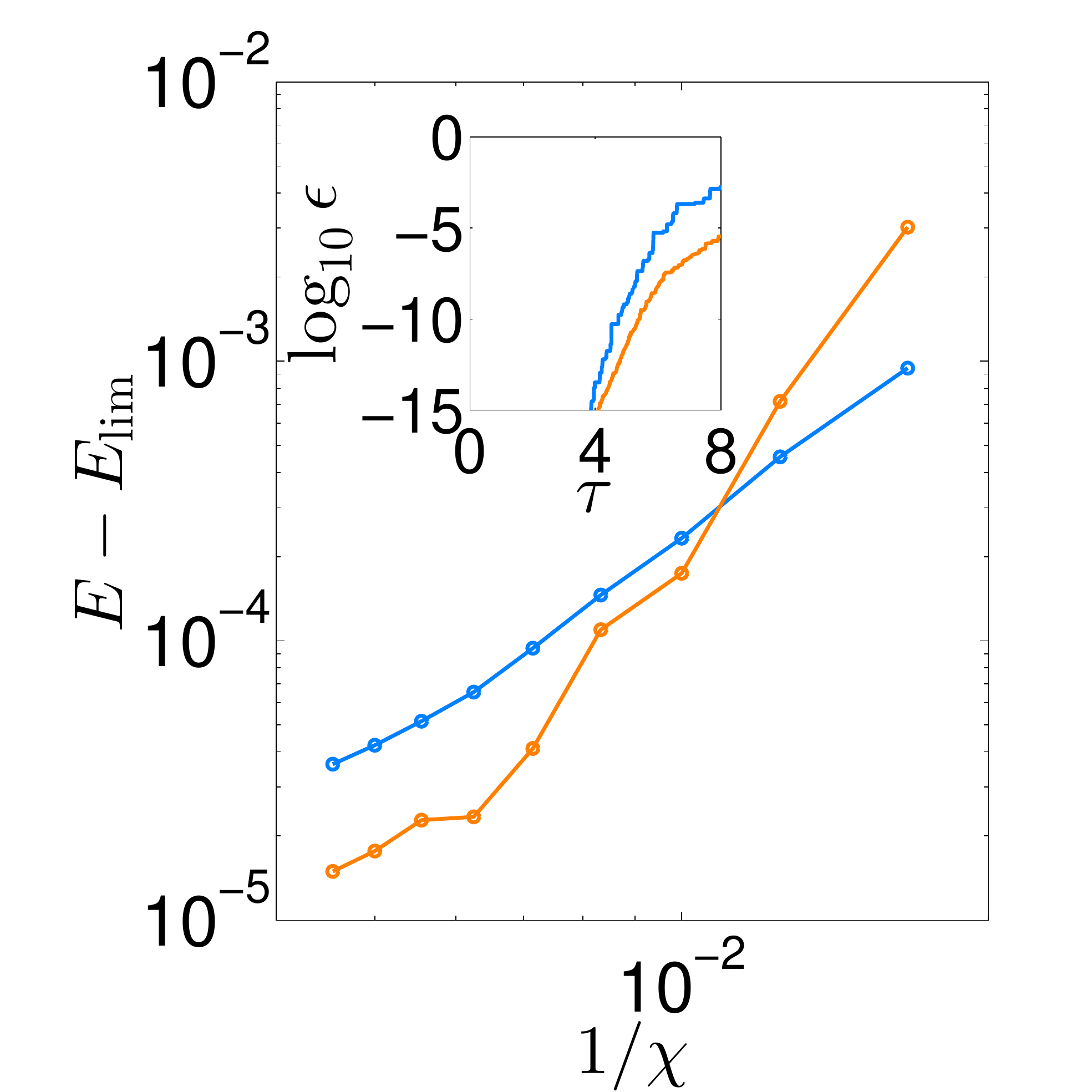,width=4.25cm,angle=0,clip=10}
\epsfig{file=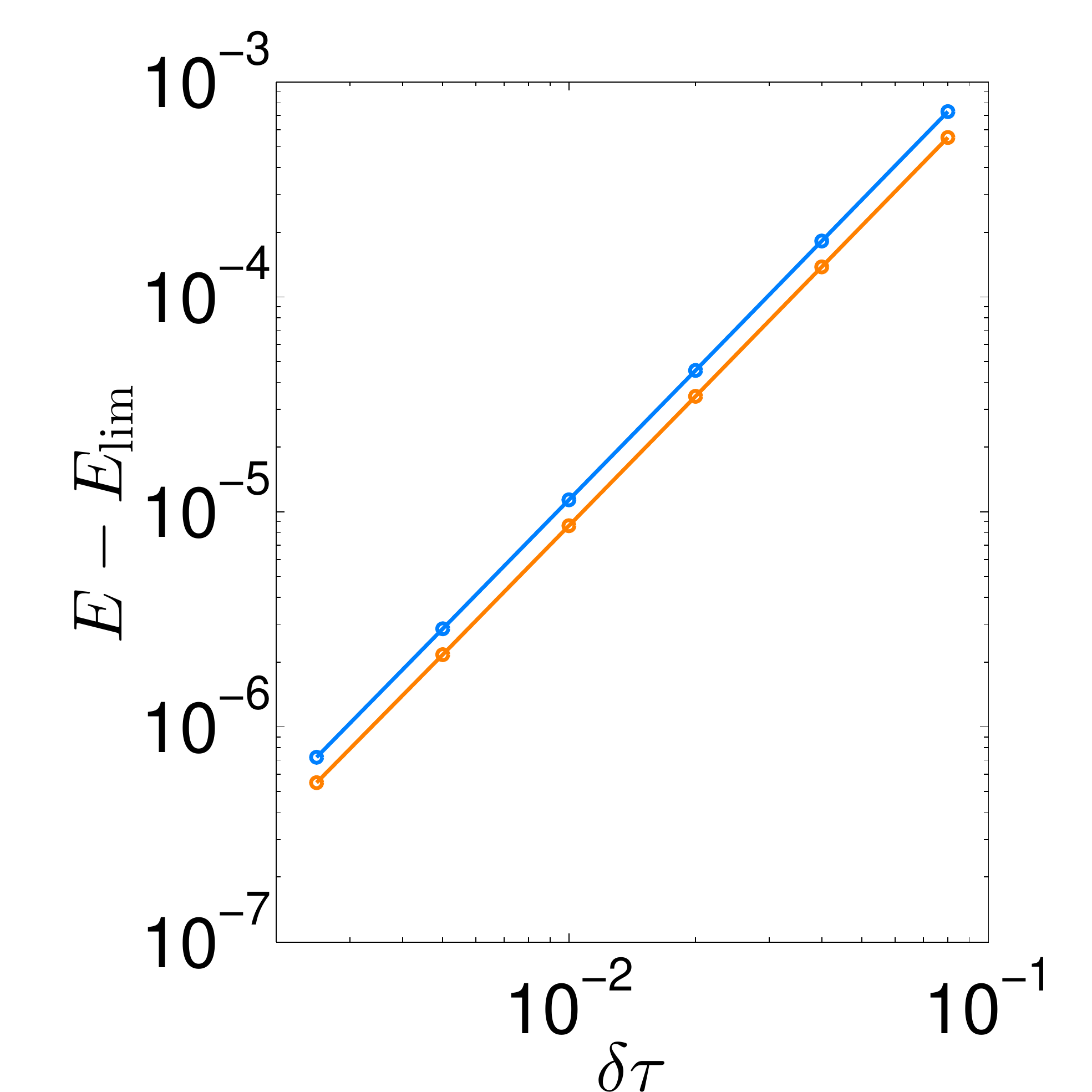,width=4.25cm,angle=0,clip=10}
\caption{Convergence tests of the electric field $E$ for the cases $m=0$, $g=0$ (blue) and $m=0.25$, $g=1.00$ (orange). Left panel: Difference of the computed electric field from the offset $E-E_{\text{lim}}$ at $\tau=8$ as a function of the inverse bond dimension $\chi$.The offset is obtained via a fit according to $E\propto (1/\chi)^b+E_{\mathrm{lim}}$.  Inset: Truncation error during the time evolution for $\chi=200$. 
Right panel: Difference of the computed electric field from the offset $E-E_{\text{lim}}$ at $\tau=2$ as a function of the time step $\delta t$. 
The offset electric field is obtained via a fit according to $E\propto \delta t^b+E_{\mathrm{lim}}$.
The exponent $b$ is in perfect agreement with the expected value of $2$.}
\label{Convergence:fig}
\end{figure}

The main sources of numerical error in our calculations are the finite bond dimension of the MPS-state representation and the Suzuki-Trotter time-step. As a bond dimension we used a value up to $\chi=200$, which ensures a truncation error on the corresponding wave function of maximum order $10^{-8}$ and $10^{-3}$ for $\tau=5$ and $9$ respectively for the string breaking calculations, and $10^{-5}$ for $\tau = 250$ for the scattering processes. Examples of the change of the truncation error during the evolution can be seen in the inset of the left panel in Fig. \ref{Convergence:fig}.

In order to make sure that these errors lead to small changes in our main observables, we tested the convergence of the mean electric field in the center of the chain (cf. Fig. \ref{StringBreakingEvolution:fig}) for different values of $\chi$ and typical system parameters. The results can be seen in the left panel of Fig. \ref{Convergence:fig} where we plot the mean electric field at the end of the evolution ($\tau=8$) as a function of the inverse bond dimension $1/\chi$. The mean electric field was subtracted by a fitted offset $E_{\mathrm{lim}}$ to allow a better comparison between the two sets of system parameters.
As we see from the plot, even for rather small bond dimensions $\chi<100$ the change to the largest bond dimension used ($\chi_{\mathrm{max}}=220$) is in the order of $0.1-0.01\%$. For the bond dimension used in our simulations ($\chi=200$) the difference to the extrapolated correct value is in the order of $E-E_{\mathrm{lim}}\sim 10^{-5}$.\\
As said before, we used a second order Suzuki-Trotter decomposition with a time-step of $\delta \tau=0.01$ to simulate the time-evolution. In the right panel of Fig \ref{Convergence:fig} we report the convergence test obtained repeating the same simulation with different time-steps. As expected, we find a clear $E-E_{\mathrm{lim}}\sim \delta\tau^2$ dependence: This clean power-law behavior allows us to give a very good estimate of the correct value of $E_{\mathrm{lim}}$. In summary, the error from the Suzuki-Trotter decomposition using a time-step of $\delta \tau=0.01$ is of the order of $E-E_{\mathrm{lim}}\sim 10^{-5}$, that is, of the same order as the error introduced by the truncated bond dimension.

\begin{figure}
\epsfig{file=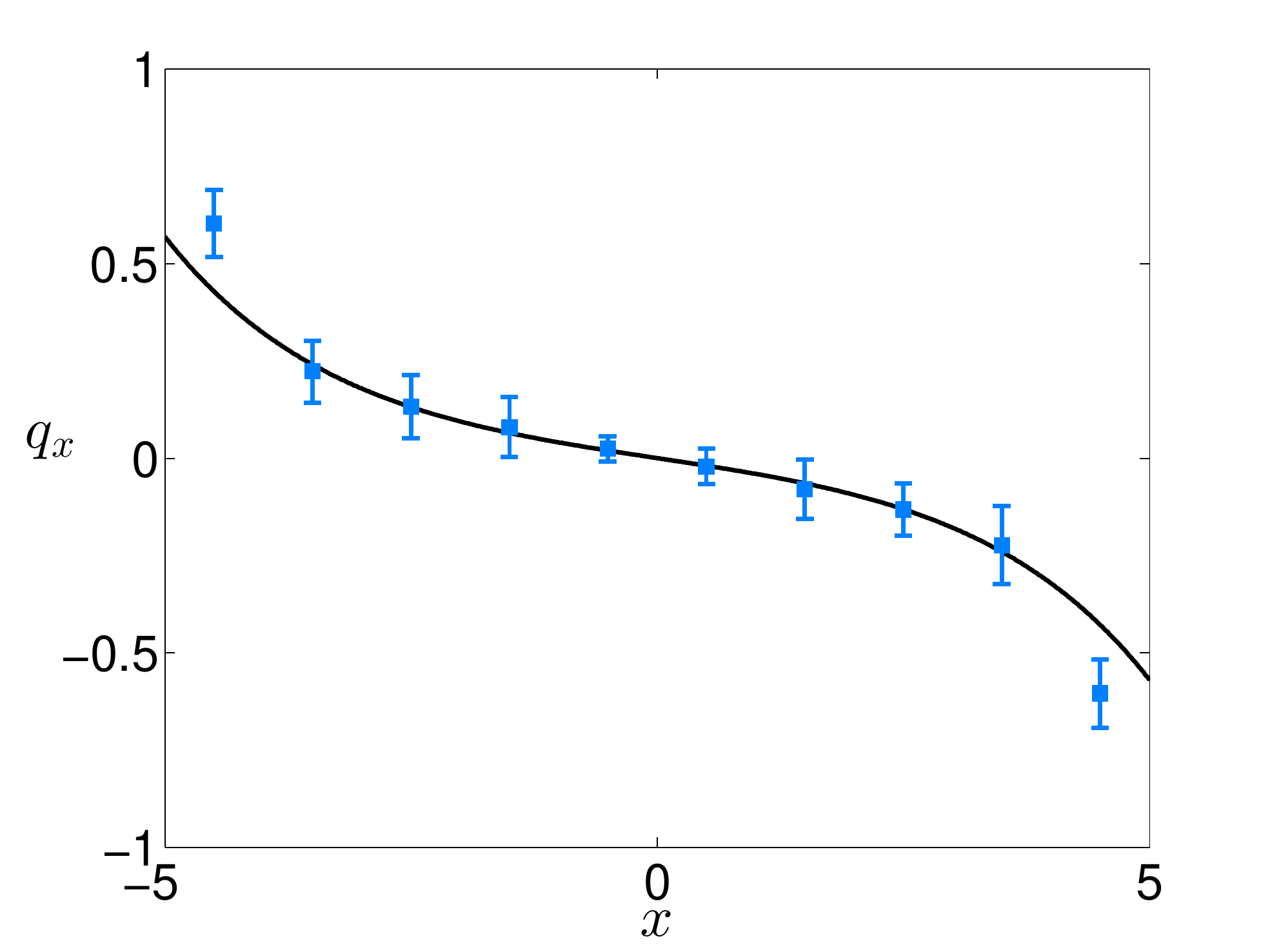,width=8cm,angle=0,clip=10}
\caption{Screening of external static charges after a time $\tau=400$ using $m=0.05$, $g=1$, and $\chi=80$  (blue squares). Two numerical lattice sites are combined to form one physical lattice site due to our staggered fermions formulation. The dynamical charge density decays exponentially with the distance from the charges, the black line represents the theoretical model with $q(x) = a\exp(-bx)-a\exp(bx)$ where the decay parameter is $b=g/\sqrt{\pi}$ and the scale parameter is $a=\exp(b)-1$.}
\label{Screening:fig}
\end{figure}

\section{Screening of Charges}
\label{AppB}
From the earliest discussion of the Schwinger model, the screening of static charges due to vacuum polarization has attracted a lot of interest \cite{Schwinger2,Casher}. Recent semiclassical calculations have reproduced this process including also finite masses where it could be shown that the main dynamics for $m<g$ is mostly independent from the mass~\cite{HebenstreitPRD}.
Here, we show that it is possible to study such process using quantum link models implemented in a tensor network algorithm, already with very small spin representation, e.g. $S=1$. As the screening of charges can be  observed only in asymptotically large times, the large correlations building up during the process have carefully to be kept under control, otherwise they would undermine the efficiency of our approach. 
Thus, here we limit our system size to $20$ lattice sites and encode the two static charges which build up the initial string in the boundary conditions on both sides of the lattice. In such a way, we avoid to simulate the 
vacuum regions where quantum correlations build up very quickly (see Fig.~\ref{Overview:fig}) and are less interesting for the phenomena under study. 
Thus, we simulate the time evolution of the string of electric field among the charges with a small, but finite mass:  after a long time compared to other timescales of the system ($\tau=400$) when the largest fluctuations in the lattice damped out, and report the final charge distribution $ q_x $ (averaged over the last $\tau=10$ to remove remaining oscillatory effects) in Fig.~\ref{Screening:fig}. 
The net results is an exponential decay of the charge density with the distance from the charges, as discussed in Ref. \cite{HebenstreitPRD} where the authors showed that the screening also at $m \ne 0$ is equal to that experienced by the massless Schwinger model, as $ q_x  = a\exp(-bx)$ with a decay rate $b=g/\sqrt{\pi} \approx 0.5642$~\cite{Schwinger2}. The parameter $a=\exp(b)-1$ is derived from the normalization condition $\sum_{x=1}^{\infty} q_x=1$.
It can be clearly seen that our findings are compatible with the theoretical results from Ref. \cite{HebenstreitPRD} (black solid line in Fig.~\ref{Screening:fig}).

We stress that the goal of the presented analysis is to show that it is possible to perform a study of  the screening of charges using tensor networks, not to perform a thorough discussion which will be presented elsewhere.  Indeed, an extensive numerical analysis is needed to individuate or rule out possible deviations of the massive from the massless case in a full quantum mechanical model, the influence of the spin representation in the quantum link formulation, possible finite size effect, and also more challenging regimes of parameters, e.g. $m \sim 1$ as shown in~\cite{HebenstreitPRD}. However, the agreement between the presented results and the semiclassical approach, indicates that both analysis are capable of describing the physics of the system and might be alternative approaches which can complement and benchmark each other.

\end{document}